\documentclass[hyper(*)]{JHEP3} 

\usepackage{epsfig,multicol,bbm}

\newcommand\fverb{\setbox\fverbbox=\hbox\bgroup\verb}
\newcommand\fverbdo{\egroup\medskip\noindent%
            \fbox{\unhbox\fverbbox}\ }
\newcommand\fverbit{\egroup\item[\fbox{\unhbox\fverbbox}]}
\newbox\fverbbox

\title{Hitchin Equation, Singularity, and $N=2$ Superconformal Field Theories}

\author{Dimitri Nanopoulos $^{1,2,3,a}$, and Dan Xie $^{1,b}$

\\$^{1}$George P. and Cynthia W.Mitchell Institute for Fundamental Physics,
Texas A\&M University, College Station, TX 77843, \\
$^{2}$Astroparticle physics Group, Houston Advanced Research
Center (HARC), Mitchell Campus, Woodlands, TX 77381, USA\\
$^{3}$Academy of Athens, Division of Nature Sciences, 28
panepistimiou Avenue, Athens 10679, Greece\\
E-mail: \email{$^{a}$dimitri@physics.tamu.edu,
$^{b}$fogman@neo.tamu.edu}}

\preprint{ACT-11-09, MIFF-09-46}  

\abstract{We argue that Hitchin's equation determines
not only the low energy effective theory but also describes the UV theory of four
dimensional $N=2$ superconformal field theories when we compactify six dimensional
$A_N$ $(0,2)$ theory on a punctured Riemann surface. We study singular solutions to Hitchin's equation and
the Highs field of equation has a simple pole at the punctures;
We show that the massless theory is associated with Higgs field whose residue is a nilpotent element;
We identify the flavor symmetry associated with the puncture by studying the singularity of closure of the
moduli space of solutions with the appropriate boundary conditions. For mass-deformed theory the residue
of the Higgs field is a semi-simple element, we identify the semi-simple element by arguing that
the moduli space of solutions of mass-deformed theory must be a deformation of the closure of the
moduli space of  massless theory. We also study the Seiberg-Witten curve by identifying it as the spectral curve of the Hitchin's system. The results are all in agreement with
Gaiotto's results derived from studying the Seiberg-Witten curve of four dimensional quiver gauge theory.}


\dedicated{}

\begin{document}


\section{Introduction}
$D=6$ is the maximal dimension in which we can formulate a
superconformal field theory (SCFT). Six dimensional  $(0,2)$
superconformal field theory has the famous ADE classification. The
compactification of this six dimensional theory on a Riemann surface
provides a lot of insights on four dimensional conformal field
theory \cite{witten1}.  For instance, if we compactify $A_{N-1}$
theory on a smooth torus, the $SL(2,Z)$ duality invariance of four
dimensional $ N=4$ SU(N) gauge theory is directly related to
$SL(2,Z)$ modular group of the torus.

We may wonder if we can also find a six dimensional description of four
dimensional $N=2$ superconformal field theory; In analogy with
$N=4$ theory, the duality of four dimensional field theory can
be interpreted geometrically as the property of Riemann surface on which we compactify the six dimensional theory.
Motivated by earlier work on $N=2$ $S$ duality \cite{Argy}, Gaiotto \cite{Gaiotto1}
provided a six dimensional framework to understand $S$ duality
of four dimensional $N=2$ scale invariant theory. Here we need to turn on
codimension two defects \cite{witten2} of six dimensional theory.
These defects are labeled by Yang tableaux from which we can
also read the flavor symmetries of four dimensional theory.

Gauge couplings of four dimensional theories are interpreted as the complex
structure of this punctured Riemann surface. The $S$ duality of gauge theory is
realized as the conformal mapping group of the complex structure space. 
Different weakly coupled four dimensional theories are described
as the different degeneration limits of punctured Riemann surface.

There are more information encoded in  thoes punctures.
The punctures are labeled by a Yang tableaux of total boxes $N$ and we can read 
four dimensional flavor symmetry associated with it. The Seiberg-Witten curve \cite{witten4,witten5}
is also described by a subspace in the cotangent
bundle of this Riemann surface, it is entirely determined by the information encoded
in the Yang tableaux. After the
description of this idea in \cite{Gaiotto1}, there are a lot of
developments along this idea to understand $N=2$ SCFT \cite{SO,web,N1,dan2, yuji2, zhou, surface}; See also the interesting relation between
four dimensional theory and two dimensional conformal field theory \cite{Gaiotto2,toda,wilson,wilson1,wilson2,
morozov1,mich1,morozov 2,mich2,morozov3,dan1,Gaiotto3,Gaiotto4,vafa,stony,mich3,morozov 4,morozov 5,morozov 6,
morozov 7,morozov 8}.

Gaiotto discovered above description by using  brane construction
\cite{witten3} and explored the Seiberg-Witten curve. It is important to
know whether we can have a truly six dimensional description, namely,
we want to know what is the dynamical equation living on  Riemann
surface when we compactify six dimensional theory on a Riemann surface.
In this paper, we argue that it is  Hitchin's equation which governs
the dynamics of this compactification.

Hitchin's equation has appeared before as a description of certain $N=2$ low energy
effective theory \cite{donagi}. That is based on observation about the relation between
Seiberg-Witten theory and integrable system. In this paper, we
will show that Hitchin's equation not only describes low energy
effective theory but also describes the four dimensional UV theory for a large class
of $N=2$ SCFT. We study singular solution
to Hitchin's equation and the Higgs field of solution has simple pole at singularity;
We show that the massless theory is associated with the Higgs field whose residue is a nilpotent element;
We identify the flavor symmetry associated with the puncture by studying moduli space
of solutions with appropriate boundary conditions. For mass-deformed theory the residue
of the Higgs field is a semi-simple element, we identify the semi-simple element by arguing that
the moduli space of solutions of mass-deformed theory must be a deformation of closure
of the moduli space of massless theory. We also study the Seiberg-Witten curve by identifying it as the spectral curve of the
Hitchin's system. The results are all in agreement with
Gaiotto's results derived from studying the Seiberg-Witten curve of four dimensional quiver gauge theory.

This paper is organized as follows: In section 2, we review the connection between
six dimensional theory $(0,2)$ theory and four dimensional $N=2$ superconformal
field theory. In section 3, we study string duality of certain brane configuration engineering $N=2$ theory and
argue that Hitchin's equation can be used to study a large class $N=2$ SCFT; In section 4,
we study Hitchin's equation for $SU(2)$ gauge group and prove that it is
the correct description of IR and UV behavior of four dimensional $N=2$ $SU(2)$ quiver theory;
In section 5, we generalize $SU(2)$ results to $SU(n)$ case; In section 6, we give the
conclusion and discuss some future directions. In appendix I, we give some mathematical introduction
to Nahm's equation and discuss the isomorphism between the moduli space of solutions 
and the adjoint orbit of $SL_n$ lie algebra.

\section{Six dimensional $(0,2)$ $A_N$ theory and Four dimensional $N=2$ SCFT}
We can construct a large class of four dimensional $N=2$ superconformal field
theories by using Type IIA brane configurations. The NS5 branes
which extend in the direction $x^0, x^1,x^2,x^3,x^4,x^5$, are
sitting at $x^7,x^8,x^9=0$ and at the arbitrary value of $x^6$. The
$x^6$ position is only well defined classically. The D4 branes are
stretched between the fivebranes and their world volume is in $x^0,x^1,x^2,x^3$ direction;
These D4 branes have finite length in $x^6$ direction. We also have $D6$ branes which
extend in the direction $x^0,x^1,x^2,x^3,x^7,x^8,x^9$. Two typical
brane configurations are depicted in Figure 1.
\begin{figure}
\begin{center}
\includegraphics[width=3.5in,]
{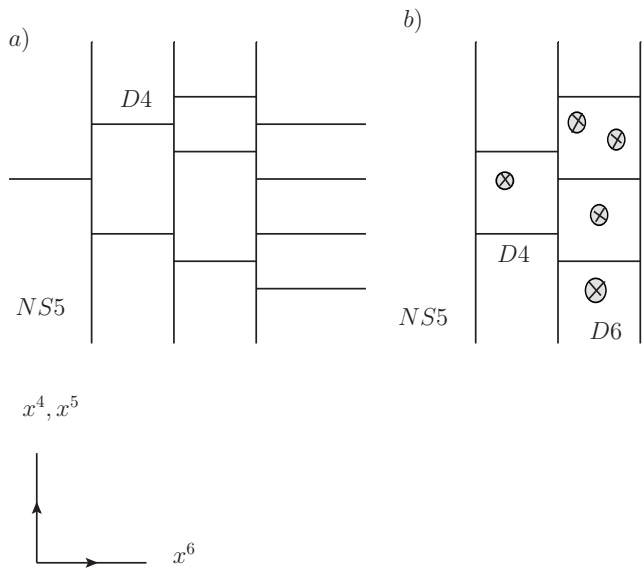}
\end{center}
\caption{ (a) A Type IIA NS5-D4 brane configuration which gives four dimensional
$N=2$ superconformal field theory, there are semi-infinite D4 branes
on both ends which provide the fundamental hypermultiplets; (b) We use $D6$ branes instead
of semi-infinite D4 branes to provide the fundamental hypermultiplets.}
\end{figure}

There are two different ways to introduce fundamental hypermultiplets to the
gauge groups at both ends: we can either attach semi-infinite D4 branes as in Figure 1a) or
add D6 branes as in Figure 1b). In this section, we only consider brane configurations with
$D6$ branes. Let's consider a brane configuration with $n+1$ $NS5$ branes and a total of $k_\alpha$ D4
branes stretched between $\alpha$th and $(\alpha+1)$th $NS5$ brane, the gauge group is 
$\prod_{\alpha=1}^nSU(k_\alpha)$, and there are bifundamental hypermultiplets transforming
in the representation $(k_\alpha,\bar{k}_{\alpha+1})$; To make the theory conformal, we need to add $d_\alpha$ fundamental
hypermultiplets to $SU(k_\alpha)$ gauge group using $D6$ branes. $d_\alpha$ is given by
\begin{equation}
d_\alpha=2k_\alpha-k_{\alpha+1}-k_{\alpha-1},
\end{equation}
where we understand that $k_0=k_{n+1}=0$.

The Seiberg-Witten curve for this theory is derived by lifting the above configuration to M theory. The $D6$ branes
are described by Taub-NUT space. $NS5-D4$ brane configurations become a single M5 brane embedded in
D6 branes background. Define coordinate $v=x^4+ix^5$ and polynomials:
\begin{equation}
J_s=\prod_{a=i_{s-1}+1}^{i_s}(v-e_a),
\end{equation}
where $1\leq s \leq n$ and $d_\alpha=i_\alpha-i_{\alpha-1}$, $e_a$ is the constant
which represents the position of D6 brane. The Seiberg-Witten curve is
\begin{eqnarray}
y^{n+1}+g_1(v)y^n+g_2(v)J_1(v)y^{n-1}+g_3(v)J_1(v)^2J_2(v)y^{n-2}\nonumber\\
+...+g_\alpha\prod_{s=1}^{\alpha-1}J_s^{\alpha-s}y^{n+1-\alpha}+...+f\prod_{s=1}^nJ_s^{n+1-s}=0 \label{SW},
\end{eqnarray}
here $g_\alpha$ is a degree $k_\alpha$ polynomial of variable $v$. The Seiberg-Witten differential
 is given by $\lambda={vdt\over t}$.

The gauge couplings are determined by $x^6$ positions
of the NS5 branes. If the beta functions for all the gauge groups vanish, the asymptotic
behaviors of the roots of Seiberg-Witten curve regarded as a polynomial in $y$ determine the gauge couplings.
In large $v$ limit, the roots are $y\sim \lambda_iv^{k_1}$,
where  $\lambda_i$ are the roots of the polynomial equation:
\begin{equation}
x^{n+1}+h_1x^n+h_2x^{n-1}+....+h_nx+f=0.
\end{equation}
In $x$ plane, there are $n+3$ distinguished points, namely $0,\infty,$ and $\lambda_i$. The
choice of $\lambda_i$ determines the asymptotic distances in coordinate $x^6$ between fivebranes
and hence the gauge
coupling constants. The
gauge coupling space is the complex structure moduli of the sphere with $n+3$ marked points among which
$0,\infty$ are distinguished. Denote the moduli space as $M_{0,n+3;2}$,
the fundamental group $\pi_1(M_{0,n+3;2})$ is interpreted as the duality group.

There are two Riemann surfaces in describing the theory: one is used to determine bare gauge couplings and the other determine the low
energy effective theory, see Figure 2 for an illustration. Is there any connection between them? Another question is: Can we get four dimensional theory by compactifying certain higher dimensional theory on this Riemann surface with marked points so that $S$ duality is manifest?
\begin{figure}
\begin{center}
\includegraphics[width=3.5in,]
{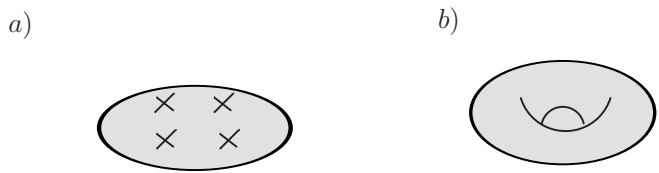}
\end{center}
\caption{ (a) Riemann surface with punctures whose complex structure moduli determines the four
dimensional gauge couplings, we take SU(2) with four fundamentals as an example; (b) Seiberg-Witten curve which determines the low energy effective action, in fact
we have a family of these curves which are parameterized by Coulomb branch parameters, it may be degenerate for 
certain parameter.}
\end{figure}

Gaiotto \cite{Gaiotto1} found the connection between these two Riemann surfaces by transforming the Seiberg-Witten
curve to the following form
\begin{equation}
x^n=\sum_{i=2}^n\phi_i(z)x^{n-i},
\end{equation}
here $\phi_i(z)$ is a degree $i$ meromorphic differential on the punctured Riemann surface which
we use to determine the gauge couplings. $\phi_i(z)$ has poles at those punctures $0,\infty,\lambda_i$.
The parameters of these degree $i$ differential are interpreted as four dimensional field theory
operators with scaling dimension $i$. The Seiberg-Witten differential is $\lambda=zdx$. Put it
in another way, the Seiberg-Witten curve is a subspace in the cotangent bundle $T^*\Sigma$, where
$\Sigma$ is our punctured Riemann surface.

To answer the second question, Gaiotto proposed that four dimensional $N=2$ SCFT
can be derived by compactifying six dimensional $(0,2)$ $A_{N-1}$ theory on this punctured Riemann surface.
The gauge coupling constants of four dimensional theory only depend on the complex 
structure of Riemann surface, this means that
the $S$ duality group is identified with the conformal mapping group of complex structure space. We need to turn on
defects on the punctures when we do the compactification.  With this interpretation, all 
information about four dimensional theory is encoded in this punctured Riemann surface. In particular,
the global symmetry is encoded in the description of the puncture. Gaiotto showed
that the punctures are labeled by a partition of $N$ and can be described by the Young
tableaux. A example is shown in Figure 3. If the Young tableaux has $l_h$ columns with height $h$, then
the flavor symmetry associated with this puncture is
\begin{equation}
S(\sum_hU(l_h)).
\end{equation}
\begin{figure}
\begin{center}
\includegraphics[width=2.5in,]
{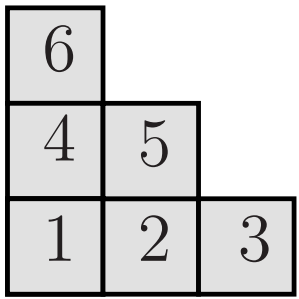}
\end{center}
\caption{A Young tableaux with total of boxes 6, the order of poles of the
meromorphic differential $\phi_i$ are: $p_1=1-1=0,p_2=2-1=1,p_3=3-1=2,p_4=4-2=2,p_5=5-2=3,p_6=6-3=3$; the
flavor symmetry associated with this puncture is U(1). }
\end{figure}
It is remarkable that this same Young tableaux determine the Seiberg-Witten curve as well. The order
of pole of $\phi_i$ at the puncture labeled by a Young tableaux is $p_i=i-s_i$, where $s_i$ is the height
of $i$th box in the Young tableaux, the number of parameters of this differential can be calculated by
using Riemann-Roch theorem.

The above results are mainly derived by studying the Seiberg-Witten curve. It is really interesting to
find a direct six dimensional description. However, there is no lagrangian description of $(0,2)$ $A_N$
theory, we really don't know what is the dynamical equation which governs the compactification. It is
the purpose of this paper to provide such a description. Hopefully, we can also get the low energy effective
theory from this equation. It will be proven in later sections that this can be done.

By realizing the four dimensional theory as compactification of six dimensional theory, $S$ duality of the
four dimensional theory is transparent since the compactification only depends on the complex structure
of the punctured Riemann surface, and four dimensional theory is invariant under the conformal mapping group
of complex structure space of the punctured surface. It is also becoming clear that the different weakly
coupled four dimensional theories at the cusp of the coupling space are realized as the different
degeneration limits of the punctured Riemann surface. We show an example of $SU(2)$ theory with four
fundamental fields in Figure 4.
\begin{figure}
\begin{center}
\includegraphics[width=4in,]
{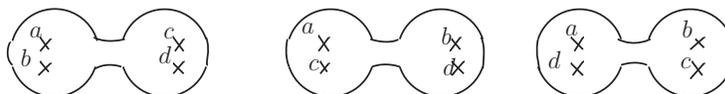}
\end{center}
\caption{The various weakly coupled limit of SU(2) theory with four
fundamental matter. The narrow strip denotes the weakly coupled
SU(2) gauge group. The punctures are associated with flavor symmetry
$SU(2)$.}
\end{figure}

We can construct more general four dimensional $N=2$ SCFT by using the sphere with three
full punctures (the flavor symmetry with this puncture is $SU(N)$). This theory has no gauge
couplings and has no lagrangian description for $N>2$; We call this theory as $T_N$, it is shown in Figure 5.
Using this build block $T_N$, we can construct a large class of  generalized quiver gauge theories which don't
have conventional lagrangian descriptions, see an example in Figure 6;
A S dual generalized quiver gauge theory of Figure 6 is shown in Figure 7, the quiver corresponds to a different
degeneration limit of genus two Riemann surface.
\begin{figure}
\begin{center}
\includegraphics[width=4in,]
{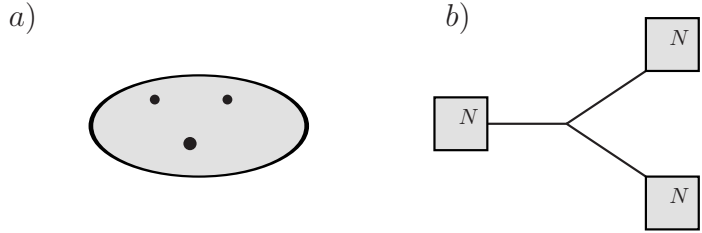}
\end{center}
\caption{a) The sphere with three full punctures; b)The graph representation of the theory derived from compactification on a).}
\end{figure}

\begin{figure}
\begin{center}
\includegraphics[width=4in,]
{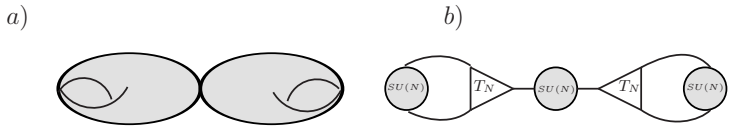}
\end{center}
\caption{a) Genus two Riemann surface without punctures and one of its degeneration limit; there are three nodes
and we identify them with gauge groups; note the nodes are equivalent with our previous representation 
by long necks; b) A generalized quiver description with a six
dimensional compactification on a).}
\end{figure}

\begin{figure}
\begin{center}
\includegraphics[width=4in,]
{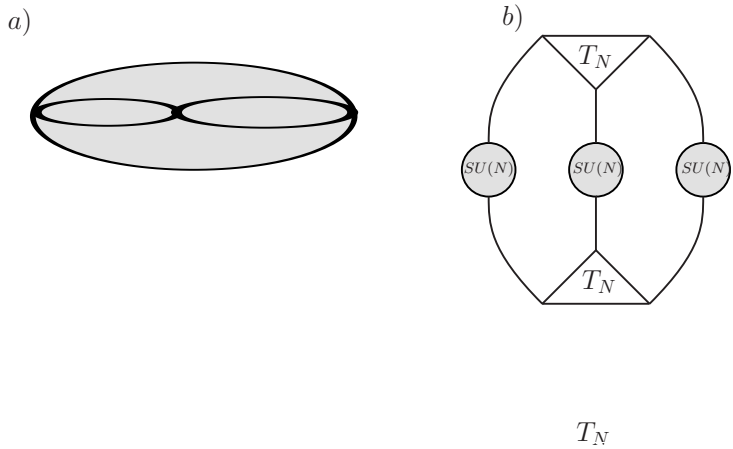}
\end{center}
\caption{a) a) Another degeneration limit of genus two Riemann surface without punctures. b) A
generalized quiver corresponding to a).}
\end{figure}

\section{Hitchin's equation }
In this section, we will use brane configuration and string dualities
to find Hitchin's equation which is used to describe the Coulomb branch
of $N=2$ SCFT.

The property of Coulomb branch of $N=2$ theory is described by
Seiberg-Witten curve. The Coulomb branch of the theory is a special Kahler
manifold of complex dimension $r$, where $r$ is the dimension of 
Cartan algebra of gauge group. The moduli space $U$ is
topologically $C^r$. The special Kahler metric is determined by a
function: prepotential. This function is multivalued and
Seiberg-Witten considered a more invariant description: Seiberg-Witten
fibration. Consider a fibration $\pi:X\rightarrow U$, where $X$ is a
complex manifold with dimension 2r, and the fibers are Abelian
varieties $A_r$ of dimension $r$(in other words, $A_r$ is a complex
Riemann surface whose first homology class has dimension 2r, and we
have a $(1,1)$ form $t$ which is positive and has integral periods).
We also need a holomorphic $(2,0)$ form $\omega$ on $X$ whose
restriction to the fibers of $\pi$ is zero.

The metric on moduli space is calculated by taking a $(r+1,r+1)$
form $t^{r+1}\wedge\omega\wedge\bar{\omega}$ and integrating it over the
fibers of $\pi$. The $\omega$ is taken nondegenerate away from the
singular fiber. We also need to specify the coupling constant of the
low energy theory; these coupling constants are determined by the
complex structure constant of the fibers.

On the other hand, we can think of $X$ as a complex symplectic
manifold which is identified with the complex phase space for a
mechanical system.  Since the restriction of $\omega$ on the fibre
is zero, this is a integrable system. The coordinates on $U$ are
seen as the coordinate, while the coordinates on the fibre are
thought of as the conjugate momentum. Thus any $N=2$ system
corresponds to a certain complex integrable system of classical
mechanics.

It is useful to make a direct connection with these two approaches. Kapustin
made this connection for a certain $N=2$ model \cite{kapustin}, we will review his
derivation below. We start with the elliptic model studied by Witten \cite{witten3}.
The brane configuration is almost the same as we described in last section
with the difference that we take $x^6$ to be compact, see Figure 8a).
The low energy field theory on this system is a four dimensional $N=2$ SCFT, it
is called elliptic model. The gauge group is
$SU(k)^{n}\times U(1)$. The solution of the low energy theory is also solved by
using M theory.
\begin{figure}
\begin{center}
\includegraphics[width=4in,]
{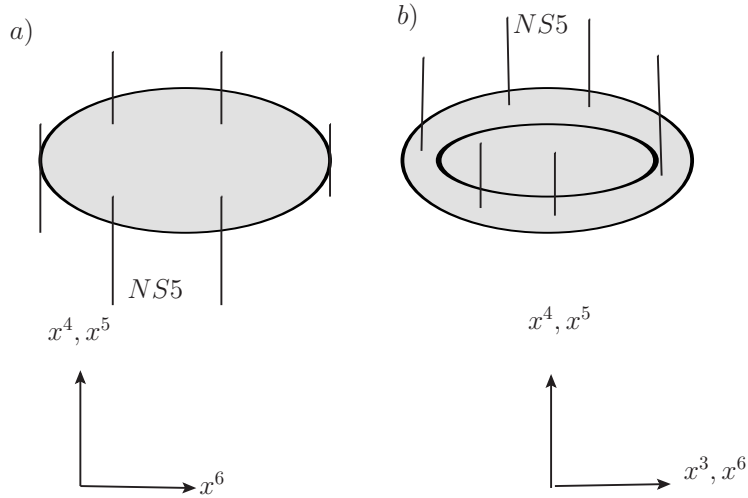}
\end{center}
\caption{a)The electric brane configuration of elliptic model; b) The
magnetic brane configuration of elliptic model.}
\end{figure}

Now we compactify further coordinate $x^3$, the theory becomes
effectively a three dimensional theory. The Coulomb branch of the
low energy theory is described as a hyperkahler manifold $X$ with a
distinguished complex structure in which it looks like a fibration $
\pi:X\rightarrow C^r$ with fibers being abelian varieties $A_r$ with
complex dimension $r$ \cite{3d}. We first do a T duality on $x^3$ and then do
a Type IIB S duality and finally another T duality on $x^3$
coordinate, We finally come back to Type IIA configuration. The NS5
brane becomes a IIB NS5 branes under first T duality; S duality turns it
into IIB D5 brane, and the second T duality turns it into IIA D4 brane
located at fixed $x^3, x^6, x^7, x^8, x^9$, we call it $D4^{'}$
brane; The original $D4$ branes are not changed. The whole brane
configuration becomes $D4-D4^{'}$ system: D4 brane wrapped on $x^3,
x^6$ torus and $D^{4'}$ is sitting at a fixed point of the torus.
The gauge theory on $D4$ branes is a $U(k)$ theory with fundamentals
coming from the string stretched between the $D4$ branes and
$D4^{'}$ branes.  The Coulomb branch of the original theory is
matched to Higgs branch of the dual theory which does not receive
quantum corrections and can be calculated classically. The theory on
D4 branes are five dimensional Super Yang-Mills (SYM) theory and $D4^{'}$
branes are codimension two impurities on this five dimensional theory.

We now make a connection with Gaiotto's description. 
The three dimensional theory in the dual
system is interpreted as five dimensional SYM theory compactified
on a torus with impurities, those impurities are coming from 
$D4^{'}$ branes. Now let's lift it to M theory. There is another compact
coordinate $x^{10}$, and the original D4 branes are M5 branes wrapped on a
three torus $T^3$. In the dual description, we have M5 branes which from D4 branes
wrapped on $x^3, x^6, x^10$ and M5 branes from $D4^{'}$ branes are sitting at a point of
the torus $x^3, x^6$. The effective theory can be interpreted as M5 branes
compactified on a three dimensional manifold:
We first compactify n M5 branes on a circle and then compactify
it on a pucntured torus, we get a three dimensional theory. Interesting things happen if we change the order of compactification,
we first compactify it on a punctured torus, and then on a circle, we can get
the same three dimensional theory. We also assume the description of the theory on punctured
torus is not changed. Now we take the circle to infinity and
get back to a four dimensional theory, and we get the
picture discovered by Gaiotto:  Four dimensional $N=2$ theory can be described
as $n$ M5 branes compactified on a punctured torus, see Figure 8b). The second series
of the compactification can be interpreted as regarding $x^3$ as the M theory cycle.

Now the Coulomb branch of the original theory is mapped to Higgs
branch of the five dimensional theory on a torus with punctures. The
moduli space of Higgs branch is described by the Hitchin's equation \cite{kapustin1}:
\begin{equation}
F_{z\bar{z}}-[\Phi_z,\Phi_z^{+}]=0
\end{equation}
\begin{equation}
\bar{D}\Phi_z={-\pi\over
RL}\sum_{\alpha=1}^k\delta^2(z-z_\alpha)diag(m_\alpha, -M,...,-M).
\end{equation}
There are source terms coming from the $D4^{'}$ impurities.
The connection between the description of Hitchin's equation in dual picture and
Seiberg-Witten curve in original picture is that the Seiberg-Witten curve is
described by the spectral curve of the moduli space to Hitchin's equation, see the early attempt
of using Hitchin's equation to derive Seiberg-Witten curve \cite{donagi}. Hitchin's eqution
is an integrable system and this answers the initial question to make a connection between
Seiberg-Witten curve and integrable system. We hope to generalize this to a large class
of $N=2$ SCFT, see \cite{moore} for extensive study for the use of the Hitchin's equation on $N=2$ wall crossing.

Now we have two different descriptions of the same four dimensional gauge theory, in the original description,
gauge groups and matter contents are
explicitly described, we can define and study all kinds of observable in this
description by using conventional field theory technics. The
dual description is realized as the six dimensional theory
compactified on punctured Riemann surface, the gauge group
description is obscure but the $S$ duality property is clear. S
duality is realized as the fundamental group of complex
structure moduli space of the punctured torus. The Seiberg-Witten curve is also easily written by
exploring the Hitchin's equation.

We can also give this kind of description to the $A_N$ type conformal quiver. We first consider
the simplest case with $n$ $SU(N)$ gauge groups. We have bifundamental
hypermulitplets between the adjacent gauge groups and we need to add  $N$ fundamental hypermultiplets
at the both ends to make the whole theory conformal, see Figure 9 for a brane configuration. This model can be derived from the elliptic
model with $n+1$ $SU(N)$ gauge group: we decouple one of the $SU(N)$ gauge
group of the elliptic model and it becomes the linear quiver we are interested in. The elliptic
model is described as the six dimensional $(0,2)$ theory compactified on a torus with $n+1$ punctures.
The decoupling of one gauge group means a complete degeneration of the torus, and we are left a sphere with five
punctures, the extra two punctures come from the degeneration of the torus. We put those
two extra punctures at $0,\infty$. This
is exactly the compact Riemann surface to describe the linear quiver,
see Figure 10 for illustration.

\begin{figure}
\begin{center}
\includegraphics[width=4in,]
{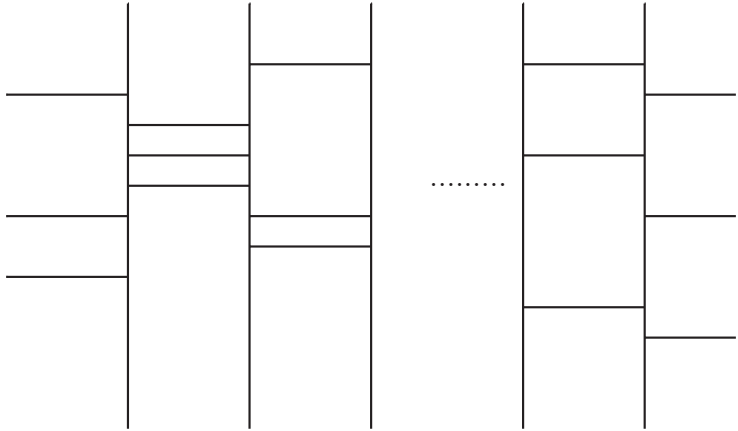}
\end{center}
\caption{The brane configuration for $N=2$, $\prod_{\alpha=1}^n SU_\alpha(3)$ superconformal field theory.}
\end{figure}

\begin{figure}
\begin{center}
\includegraphics[width=4in,]
{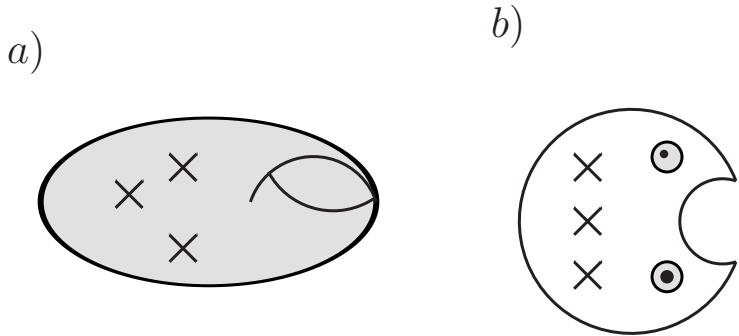}
\end{center}
\caption{a)Torus with 3 punctures; b)After degeneration, we have a sphere with 5 punctures, the
two punctures coming from the degeneration of torus are different from other punctures.}
\end{figure}
It is conceivable that Coulomb branch of the linear quiver is still described by Hitchin's equation with
the source terms coming from the punctures, the difference here is that the source terms at
$0,\infty$ are different from other punctures. The Seiberg-Witten curve
is described by the spectral curve of the Hitchin system.

This conjecture may be seen from the Seiberg-Witten curve. The linear curve is described in Figure 9, and the Seiberg-Witten
curve of the massless theory is \cite{witten3}
\begin{equation}
v^Nt^{n+1}+g_1(v)t^{n}+..+g_\alpha t^{n+1-\alpha}..+v^N=0,
\end{equation}
where $g_\alpha(v)= v^N+c_2V^{N-2}+...c_N$. It is easy to see that coefficients of $t^k$ have
the same order in $v$, so this curve can be described as the spectral curve. We can find the appropriate
boundary conditions on the puncture of the Hitchin's equation by mapping the above curve to
the spectral curve of the Hitchin system \cite{moore}. When we turn on the mass,
it can be shown similarly that there is a Hitchin system description of the Seiberg-Witten curve.

For the general linear quiver gauge theory we described in section II, it is not obvious that we
can still use Hitchin's equation and write the Seiberg-Witten curve as the spectral curve. However, we can also give
a heuristic argument that this is possible by transforming Seiberg-Witten curve. The Seiberg-Witten curve does not depend on the $x^6$ position of
the $D6$ branes so we can move the $D6$ branes to  $x^6=\infty,-\infty$.  There is
Hanany-Witten effect \cite{hanany}: when we move the $D6$ branes across the NS5 branes, $D4$ branes will be created, and the initial configuration is equivalent to a brane configuration without the $D6$ branes, the number of D4 branes are different now.
See Figure 11 for an example. The brane configuration with D6 branes  moved to infinity is exactly the same
as the linear quiver we just studied in Figure 9, so we conclude that we can still use Hitchin's equation to
describe the Coulomb branch of the general linear superconformal quiver. The boundary conditions
at the $0, \infty$ are different here, since the semi-infinite branes are attached to the D6 branes sitting at $\infty$, this might give us different boundary conditions at $0$ and $\infty$ from the ones we just studied.
\begin{figure}
\begin{center}
\includegraphics[width=4in,]
{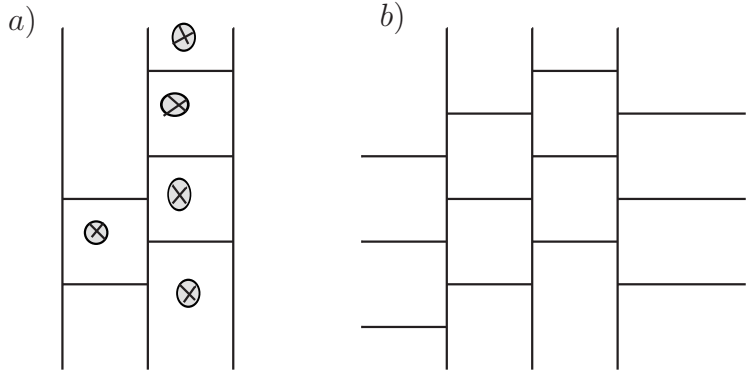}
\end{center}
\caption{a) Brane configurations with $D6$ branes sitting between the NS5 branes;
b) Equivalent brane configuration when we move all the D6 branes to the infinity.}
\end{figure}

The Hanany-Witten effect may be seen by doing some transformation on
Seiberg-Witten curve. The Seiberg-Witten curve is
\begin{eqnarray}
y^{n+1}+g_1(v)y^n+g_2(v)J_1(v)y^{n-1}+g_3(v)J_1(v)^2J_2(v)y^{n-2}\nonumber\\
+...+g_\alpha\prod_{s=1}^{\alpha-1}J_s^{\alpha-s}y^{n+1-\alpha}+...+f\prod_{s=1}^nJ_s^{n+1-s}=0 \label{SW}.
\end{eqnarray}
We study the massless theory for simplicity, so we put all the $D6$ branes at $v=0$. The number of
fundamental hypermultiplets are given by $d_\alpha=2k_\alpha-k_{\alpha-1}-k_{\alpha+1}\geq 0$, and
we have the following relation on the rank of the gauge group:
\begin{equation}
k_1\leq k_2...=k_r=...k_s\geq k_{s+1}...\geq k_n,
\end{equation}
we define $N=k_r=...k_s$. The total number of D6 branes is given by $\sum_{\alpha}d_\alpha=k_n+k_1$. We need
to redefine the $y$ coordinate so that the coefficient of $y^{'n+1}$ is the same as the constant term. Define
$y=v^ay^{'}$, then the coefficient of $y^{'n+1}$ is $v^{an+a}$. Examine the constant term:
\begin{equation}
\prod_{s=1}^nJ_s^{n+1-s}=v^{d_1n}v^{d_2(n-1)}..v^{d_n}=v^{(\sum_{i=1}^nd_i)n-\sum_{i=2}^n(i-1)d_i}=v^{(k_1+k_n)n+(k_1-nk_n)}
=v^{nk_1+k_1}.
\end{equation}
$\sum_{i=1}^nd_i=k_1+k_n$ and $\sum_{i=2}^{n}(i-1)d_i=-k_1+nk_n$ are used. We conclude that
$a=k_1$. Substituting $y=y^{'}v^{k_1}$, the coefficient before $y^{'n+1-\alpha}$ is
\begin{equation}
c_\alpha=g_\alpha(v)v^{d_1(\alpha-1)}v^{d_2(\alpha-2)}...v^{d_{(\alpha-1)}}v^{k_1n+k_1-k_1\alpha}.
\end{equation}
Calculating the exponent carefully, one finds
\begin{equation}
c_\alpha=g_\alpha v^{(n+1)k_1-k_\alpha}.
\end{equation}
Recall that $g_\alpha$ is degree $k_\alpha$ polynomial in $v$, this means that all $c_\alpha$
has the same order $(n+1)k_1$, we can have a spectral curve description!
The maximal value of $k_\alpha$ is $N$, we can factorize out $v^{(n+1)k_1-N}$ for each coefficient and we are left with
a Seiberg-Witten curve with the form:
\begin{equation}
v^Ny^{'n+1}+\sum_{\alpha=1}^{n}g^{'}_{\alpha}(v)y^{n+1-\alpha}+v^N=0,
\end{equation}
where $g^{'}_{\alpha}$ is a degree N polynomial in $v$, this is exactly the same form as the Seiberg-Witten curve of
the linear quiver shown in Figure 9, so we
can use Hitchin's equation to describe the general linear superconformal quiver!

There is a more general way in which we can see the emergence of the Hitchin's equation.
Consider six dimensional $(0,2)$ $A_{N-1}$ SCFT compactified on a punctured Riemann surface $\Sigma$ (the
following analysis is also true for a Riemann surface without singularity),
we get a four dimensional $N=2$ SCFT. To preserve some supersymmetry on the curved manifold, we actually 
need to twist the six dimensional theory \cite{Gaiotto1}. We can further compactify 
four dimensional theory on a torus $T^2$, and the two dimensional theory is a 
sigma model. The first step of compactification is hard to study since there
is no lagrangian description of the six dimensional theory. Things become clear
if we do the compactification in reverse order in the same spirit as 
we did in previous analysis on $D^4$ brane system. We first compactify on $T^2$, 
and then further on a punctured Riemann surface down to two dimension. We obtain
four dimensional $N=4$ theory with gauge group $SU(N)$ when we compacfity the theory on $T^2$, 
we then compatify the theory on a two dimensional Riemann surface $\Sigma$ and get a sigma model on two dimension. To preserve some supersymmetry,
we also need to twist $N=4$ theory, there are different kinds of twist we can make. 
The twist which is relevant for our purpose is the so called GL twist \cite{GL}, it turns
out that Hitchin's equation is the equation for the BPS condition. The theory can
also be extended to the case that the fields have singularities \cite{surface, gukov, wild}. In the language of 
Geometric Langlands program, the case with simple pole is called tame ramification 
and the case with higher order pole is called wild ramification. By comparing two
different kinds of compactification, we may conjecture that Hitchin's equation is 
the BPS equation governing the compactification of six dimensional theory on a Riemann
surface (with or without singularities). In this paper, we only consider the Hitchin's
equation without singularity and with simple singularity and leave the wild
singularity for future analysis.

\section{$A_1$ theory}
In last section, we conjecture that it is Hitchin's equation which is relevant 
when we compactify six dimensional $(0,2)$ theory on a punctured Riemann surface.
We will show in this section that Hitchin's equation provides an description of both 
UV theory and IR theory for four dimensional SCFT.

We first analyze four dimensional $N=2$ $SU(2)$ SCFT. These theories
can be described as the six dimensional $(0,2)$ $A_1$ theories compactified
on punctured Riemann surface. As we described in section 2, Gaiotto proposed
that the Seiberg-Witten curve has the form
\begin{equation}
x^2=\phi_2(z),\label{gen}
\end{equation}
where $z$ is the coordinate on  punctured Riemann surface. For massless theory,
$\phi_2(z)$ has simple pole at various punctures, and near the puncture, say $z=0$,
\begin{equation}
\phi_2(z)={c\over z}.\label{SU(2)}
\end{equation}
The Seiberg-Witten curve for massive theory is also of the form (\ref{gen}),
but the degree $2$ differential near the puncture now takes the form \cite{Gaiotto1}:
\begin{equation}
\phi_2(z)={q\over z^2}+{c\over z}.
\end{equation}

As we discussed in last section, the four
dimensional gauge theory is controlled by Hitchin equation on a
Riemann surface with possible source terms. In order
to understand the four dimensional $N=2$ SCFT, we need to solve the Hitchin's equation
with the added source terms. It turns out that for $N=2$ SCFT we reviewed in
section $2,3$, the fields should have the simple singularities at various punctures.

\subsection{Singular Solutions to Hitchin's Equation}
In previous sections, we see from brane construction that there
are added source terms on the right-hand side of Hitchin's equation due
to other branes intersecting at the puncture; These source terms
induce the singularity to the solution of Hitchin's equation. An alterative point of view is to study
the singular solution of the Hitchin's equation without the source
terms, and this will tell us enough information about the
singularity. The same problem has been studied
by Witten and Gukov in the physics approach to Geometric Langlands problem for the tame ramification case (see extensive study in \cite{witten2,gukov}), we give a review below for the necessary information we need.

The Hitchin's equation for $SU(2)$ gauge group \cite{hitchin1,hitchin2} is
\begin{eqnarray}
F_A-\phi\wedge\phi=0 \nonumber\\
d_A\phi=0,~~d_A*\phi=0,
\end{eqnarray}
here $A$ is a connection for a $SU(2)$ bundle on Riemann surface and
$\phi$ is a one form taking value on adjoint bundle and we call it Higgs field; $d_A$ is
the familiar covariant derivative. We first consider the solution without
any singularity, the moduli space of  solutions is a hyper-kahler
manifold with three complex structures $I,J,K$, and it has a
hyper-kahler quotient description. In complex structure I,
a solution of Hitchin's equation on a Riemann surface describes a Higgs bundle,
that is a pair $(E,\phi)$, where $E$ is a holomorphic G-bundle and $\phi$ is a
holomorphic section of $K_C\bigotimes ad(E)$ (here  $K_C$ is the
canonical bundle on $C$). The Higgs bundle is constructed as
follows: we interprets the $(0,1)$ part of the covariant derivative
$d_A$ as a $\bar{\partial}_A$ operator that gives the bundle E a
holomorphic structure. We denote $\Phi$ as the $(1,0)$ part of the
Higgs field $\phi$, and Hitchin's equation implies that $\phi$ is
holomorphic.

If we study the solution of Hitchin's equation with singularity at the
origin, the moduli space of solutions still has hyper-kahler structure.
In complex structure I, the solution describes a Higgs bundle but
the Higgs field has a pole at the origin. We only consider local behavior
of the solution and leave the global property for future study.
We choose local holomorphic coordinate $z=re^{i\theta}$ around the singularity.
 We only consider regular singularity here
(the solution has simple pole at $r=0$), irregular singularity is important
when we study asymptotically free theory. We consider the superconformal theory,
so we need to find conformal invariant solutions to Hitchin's equation. The most general
scale-invariant and rotation-invariant solution is
\begin{eqnarray}
A=a(r)d\theta+f(r){dr\over r} \nonumber\\
\phi=b(r){dr\over r}-c(r)d\theta.
\end{eqnarray}
$f(r)$ can be set to zero by a gauge transformation and after
introducing a new variable $s=-\ln r$, Hitchin's equation becomes
Nahm's equations:
\begin{eqnarray}
{da\over ds}=[b,c]\nonumber\\
{db\over ds}=[c,a]\nonumber\\
{dc\over ds}=[a,b].
\end{eqnarray}

The most general conformal invariant solutions are derived by setting
$a,b,c$ to constant $\alpha,\beta,\gamma$ of the Lie algebra of SU(2), and
they must commute and we can conjugate them to lie algebra of a
maximal torus of $SU(2)$. The resulting solution is
\begin{eqnarray}
A=\alpha d\theta+...\nonumber\\
\phi=\beta{dr\over r}-\gamma d\theta+... \label{massive}.
\end{eqnarray}
We ignore possible terms which are less singular than the terms
presented above.

We also want to know the behavior of the solution when we take $\alpha,
\beta, \gamma \rightarrow 0$. One may think that there is no singularity at all.
This is not the case if we note that we may have less singular terms to
the equation. When $\alpha, \beta, \gamma \rightarrow 0$, those less
singular terms play dominant role.

Indeed, we do have less singular solution to Hitchin's equation, the
Nahm's equations can be solved by:
\begin{equation}
a=-{t_1\over s+{1/f}}, ~~b=-{t_2\over s+1/f}, ~~c=-{t_3\over
s+{1/f}}, \label{massless}
\end{equation}
where $s=-\ln r$ and $[t_1,t_2]=t_3$ and cyclic permutation thereof
which are the usual commutation relations for $SU(2)$ lie algebra. A
convenient basis for $SU(2)$ is
\begin{equation}
e_1=\left(\begin{array}{cc} -{i\over2}&~~0\\
0&~~{i\over2}\end{array}\right),
~e_2=\left(\begin{array}{cc}0&~~{i\over2}\\{i\over2}&~~0\end{array}\right),
~e_3=\left(\begin{array}{cc}0&~~-{1\over2}\\
{1\over2}&~~0\end{array}\right).
\end{equation}

The choice of $f$ spoils conformal invariance, but it is not natural
to make a choice, since then the derivative of A and $\phi$ with
respect to f is square-integrable. So this solution with $f$ allowed
to fluctuate is conformal invariant. The advantage of including this
parameter is that when $f=\infty$, we get the trivial solution.
Combined the previous discussion, the second type of solution can
be thought of the zero limit of the first type.

The fact that the second type of solution is a limit of the first
type of solution can also be seen by studying moduli space of Hitchin's equation
in complex structure $J$. In complex structure $J$, solution of Hitchin's equation describes
a flat $SL(2,C)$ bundle. It is important to study the monodromy of the
flat connection. Define complex-valued flat connection ${\cal A}=A+i\phi$
taking value in $SL(2,C)$, the monodromy is
\begin{equation}
U=P\exp(-\int_l {\cal A}),
\end{equation}
where $l$ is the contour surrounding the singularity. This monodromy
characterizes the singular behavior of the solution. The curvature
${\cal F}$ defined as ${\cal F}=d{\cal A}+{\cal A}\wedge{\cal A}$
is equal to zero due to Hitchin's equation, so the monodromy calculates as
above is independent of the contour we choose. Define
$\zeta=\alpha-i\gamma$, the
monodromy for our first set of solutions (\ref{massive}) is
\begin{equation}
U=\exp(-2\pi\zeta).
\end{equation}

The monodromy of another solution (\ref{massless}) is
\begin{equation}
U^{'}=\exp(-2\pi(t_1-it_3)/(s_1+1/f)).
\end{equation}
The conjugacy class of this matrix is independent of $s_1$ due to the property
that $(t_1-it_3)$ can be taken as up triangular form. We choose a basis
in which $t_2=e_1$, $t_1=e_3$, and $t_3=e_2$. 

Indeed, what's relevant is the conjugacy class for the monodromy. Let's denote 
the conjugacy class for U as $C_\zeta$, and the conjugacy class for $U^{'}$
is a union of $C_0$ and $C^{'}$, where $C_0$ is the conjugacy class for
identity and $C^{'}$ is the conjugacy class for unipotent orbit. It can be shown that 
as $\zeta\rightarrow 0$, $C_\zeta$ approaches the union of $C_0$ and $C^{'}$. This 
also indicates that the second set of solutions is a limit of the first set of 
solutions.

With this relation, we might think that the conjugate class
of $C^{'}$ is associated with  massless $N=2$ SCFT while $C_\zeta$
is associated with mass-deformed theory. When we compactify six dimensional
$A_1$ theory on punctured Riemann surface, Hitchin's equation is the
dynamical equation we need to solve, and the information of four dimensional
$N=2$ gauge theory is encoded in the solutions to Hitchin's equation. Moreover,
we will prove that we can read the flavor symmetry and tail of the quiver gauge theory
from the solutions; and the Seinerg-Witten
curve is the spectral curve of the Hitchin system. The UV
and IR information are both encoded in the same system. In the following parts of
this section, we will confirm this conjecture.

\subsection{Massless Theory and Mass deformed theory}

Some group theoretical definitions are useful for our later use. An element of a
complex Lie group is called semisimple if it can be diagonalized (or
conjugated to a maximal torus). The conjugate class of this element
is called semisimple accordingly and it is closed. This element
can be expressed as $U=\exp(\tilde{n})$, where $\tilde{n}$
is the semisimple element of the lie algebra ($\tilde{n}$ is
diagonizable).  In contrast,
an element $U^{'}$ is called unipotent if in any finite
representation, it takes the form $U^{'}=\exp(n)$, where $n$ is a
nilpotent element of the lie algebra. From now on, we will work on
 $sl(2,c)$ lie algebra.
Let's discuss the conjugacy classes of $sl_2$ lie algebra to give a
concrete idea about the concepts we just discussed. For each $X\in sl_2$,
we can form the conjugacy class ( sometimes we call them orbit)
\begin{equation}
{\cal O}_X=\{{A.X.A^{-1}| A\in GL_2}\}.
\end{equation}
Using $tr(AB)=tr(BA)$, the $GL_2$ conjugate of an element is
also in $sl_2$.

The semisimple elements are
\begin{equation}
X(\lambda)=\left(\begin{array}{cc}\lambda&~~0\\
0&~~-\lambda\end{array}\right),
\end{equation}
where $\lambda=\Lambda / (0)$, and $\Lambda=\{C/\{\lambda\sim
-\lambda\}\}$. We also have two nilpotent elements
\begin{equation}
Y_1=\left(\begin{array}{cc}0&~~1\\
0&~~0\end{array}\right)~~~~~~~ Y_2=\left(\begin{array}{cc}
0&~~0\\
0&~~0 \end{array}\right).
\end{equation}
$Y_2$ is special since it is both semisimple and nilpotent. We call $Y_1$ as
regular nilpotent element and its orbit regular nilpotent orbit. Those
nilpotent orbit may be labeled as Young tableaux as in Figure 12.
\begin{figure}
\begin{center}
\includegraphics[width=3.5in,]
{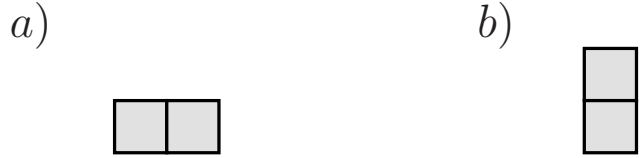}
\end{center}
\caption{ (a) Young tableaux of nilpotent element $Y_1$;
(b)Young tableaux of nilpotent element $Y_2$. }
\end{figure}

We can prove that the partition of $sl_2$ algebra has the form, see \cite{nil}:
\begin{equation}
sl_2=\bigcup_\lambda {\cal O}_{X(\lambda)}\bigcup {\cal
O}_{Y_1}\bigcup{\cal O}_{Y_2}.
\end{equation}
There are infinite number of semisimple conjugacy classes
and we only have two nilpotent conjugacy classes. We will prove that
four dimensional massless $N=2$ SCFT  theory is associated with the nilpotent
class of the lie algebra and mass-deformed theory is associated with the semisimple class.

Let's go back to Hitchin's equation. In complex structure I, the moduli space depends
on the complex structure of the Riemann surface $\Sigma$, since we want
to identify the complex structure as the gauge coupling constants, and the low energy effective
theory depends on those parameters, we will work
on complex structure I. In this complex structure, there is famous Hitchin's fibration
which we will identify as the Seiberg-Witten fibration.

The spectral curve for Hitchin system \cite{hitchin2} is
\begin{equation}
\det(x-\Phi(z))=0,
\end{equation}
where $\Phi$ is the $(1,0)$ part of the Higgs field, and the natural differential
$\lambda=xdz$ is identified with the Seiberg-Witten differential. $\Phi(z)$ is
a degree one differential on the Riemann surface, so when we expand the determinant,
the coefficient of $x^{n-i}$ is a degree $i$ differential on Riemann surface.
It is easy to
see that the characteristic polynomial only depends on the conjugate
class of $\Phi$. This spectral curve is conjectured to be the
Seiberg-Witten curve of the $N=2$ system. We will analyze the
behavior of spectral curve near the singular point $z=0$. Let's
first analyze the solution (\ref{massless})(f is not taken as
$0$), it seems that there is no simple pole for $\Phi$, since
it is less singular than $1\over r$. However, whether there is a simple
pole for $\phi$ depends on the local trivialization of the holomorphic
bundle. The $(0,1)$ part of the gauge field and $(1,0)$ part of
the Higgs field are :
\begin{equation}
A_{\bar{\omega}}d\bar{\omega}={d\bar{\omega}\over 2 (\omega+\bar{\omega})}\left(\begin{array}{cc} 1&~~0\\
0&~~-1\end{array}\right),
\end{equation}
\begin{equation}
\Phi d\omega=-i{d\omega\over \omega}{\omega\over (\omega+\bar{\omega})}\left(\begin{array}{cc}0&~~1\\
0&~~0\end{array}\right).
\end{equation}
We choose $t_1=e_1$, $t_2=e_2$, $t_3=e_3$, and $w=\ln z$ and $s=-\ln
r=-{1\over 2}(\omega+\bar{\omega})$; we take $f=\infty$ here.

The $\bar{\partial}_A$ operator is
\begin{equation}
\bar{\partial}_A=d\bar{\omega}({\partial\over\partial\bar{\omega}}+A_{\bar{\omega}})=d\bar{\omega}({\partial\over\partial\bar{\omega}}
+{1\over 2(\omega+\bar{\omega})}\left(\begin{array}{cc}
1&~~0\\0&~~-1\end{array}\right)).
\end{equation}
We can do a local trivialization to make this operator into standard form: 
\begin{equation}
\bar{\partial}_A=f\bar{\partial}f^{-1},
\end{equation}
where $\bar{\partial}=d\bar{\omega}\partial/\partial\bar{\omega}$ is
the standard $\bar{\partial}$ operator, and
\begin{equation}
f=\left(\begin{array}{cc}
({(\omega+\bar{\omega})\over\omega})^{-1/2}&~~0\\
0&({(\omega+\bar{\omega})\over\omega})^{1/2}
\end{array}\right).
\end{equation}
With this trivialization, the Higgs field becomes
\begin{equation}
f^{-1}\Phi f=-i{d\omega\over \omega}\left(\begin{array}{cc}0&~~1\\
0&~~0\end{array}\right).
\end{equation}
We conclude that the Higgs field has a simple pole at the
singularity, and its residue is proportional to the nilpotent
element. We can also add the regular constant terms to the solution, and the
full solution is
\begin{equation}
\Phi(\omega)d\omega=\left(\begin{array}{cc} a&~~{1\over\omega}+b\\
c&~~-a\end{array}\right)d\omega+{\cal \omega}(0) d\omega.
\end{equation}

Now let's analyze the spectral curve associated with this solution.
Expand the determinant, we get the equation:
\begin{equation}
x^2=Tr\Phi^2.
\end{equation}
Calculate the trace and take the singular part, we have the explicit form
\begin{equation}
x^2={c\over \omega}.
\end{equation}
This is exactly the Seiberg-Witten curve for the $SU(2)$ theory
around one of the singularity labeled by the Young tableaux in Figure 12a).
The coefficient of the pole comes from the regular term.

Consider the other solution (\ref{massive}), the holomorphic part of
the Higgs field is
\begin{equation}
\Phi={1\over2}(\beta+i\gamma){dz\over z}.
\end{equation}
We also need to include less regular terms; In proper basis, the
Higgs field has the form
\begin{equation}
\Phi=\left(\begin{array}{cc} {q\over z}&~~{1\over z}\\
0&~~-{q\over z}
\end{array}\right)dz+{\cal O}(0)dz.
\end{equation}

It can be checked that the residue of the Higgs field is conjugate with the regular semisimple element:
\begin{equation}
\left(\begin{array}{cc}
q&0\\
0&-q\end{array}\right).
\end{equation}.

Include the constant regular terms, Higgs field takes the following form:
\begin{equation}
\Phi(z)dz=\left(\begin{array}{cc} a+{q\over z}&~~{1\over z}+b\\
c&~~-{q\over z}-a
\end{array}\right)dz+{\cal O}(z)dz.
\end{equation}
The spectral curve around this singularity becomes
\begin{equation}
x^2={2q^2\over z^2} +{d\over z}.
\end{equation}
This is exactly same as the Seiberg-Wittnen curve for $SU(2)$ theory
around the singularity. The parameter for ${1\over z^2}$ term comes
from the parameter for the most singular part which is fixed, while the
parameter for ${1\over z}$ term comes from the regular term.

\subsection{Flavor Symmetry}

To understand what is the flavor symmetry associated with the massless
theory, we need to study the local singularity type of moduli space of Hithin's equation. 
The reason we study the singularity type in moduli space and identify the flavor symmetry
as the singularity type may be explained by further compactifying our four dimensional
theory to three dimensions. The moduli space of Hitchin's equation now has a physical meaning:
it is the target space of the coulomb branch of three dimensional theory \cite{3d}. The
flavor symmetry of four dimensional theory appears as the singularity type in the
coulomb branch of three dimensional theory (see also the discussion in \cite{KS}). There
are some subtleties about $U(1)$ factors though, we will discuss this
in next section since it is not relevant in this section.

Let's examine again the solution (\ref{massless}). $f$ takes values
in $R^{+}=[0,\infty)$. There is a limit for $f\rightarrow 0$, namely the trivial solution $a=b=c=0$.
We can also pick an element $R\in SO(3)$, and generalize that solution to
\begin{eqnarray}
a=-{1\over s+f^{-1}}Rt_1R^{-1}\nonumber\\
b=-{1\over s+f^{-1}}Rt_2R^{-1}\nonumber\\
c=-{1\over s+f^{-1}}Rt_3R^{-1}\nonumber.
\end{eqnarray}
So the parameter space of this family is $R^+\times SO(3)={C^2\over Z_2}$, there
is a singularity at the origin which corresponds to the trivial solution(when we
consider only the non-trivial solution, the parameter space is ${C^2/ Z_2}-\{0\}$).
It is well known that the space ${C^2/ Z_2}$ is described by the equation
\begin{equation}
a^2+bc=0.
\end{equation}
There is a $A_1$ singularity at the origin, so we identify the flavor symmetry as $SU(2)$.
In fact, this space has a nature hyper-Kahler structure. We can understand this result from group theory.
In fact, Kronheimer has found an isomorphism between the moduli space of solution to
Nahm's equation with the closure of nilpotent orbit \cite{kro1}. The interested reader can find more information
on appendix I. The nilpotent orbit of $sl(2,c)$ consists
of matrix with the condition $\det(x)=0$, write a matrix in the form
\begin{equation}
x=\left(\begin{array}{cc}
a&b\\
c&-a\end{array}\right) \label{sl2}.
\end{equation}
Calculate the determinant, we have the equation $a^2+bc=0$, this is exactly the space we found before;
we can also recognize that the identity element is at the origin of this space as we need
to set $a=b=c=0$ simultaneously. The whole space is the closure of regular nilpotent orbit.
The closure of the regular nilpotent element contains identity orbit and regular nilpotent orbit;
The identity orbit is at the boundary of the closure and is of the codimension two. The closure has
a rational singularity at the identity orbit which is at the origin of the closure.
The appearance of the codimension two singularity is a generic phenomenon for
the geometry of closure of nilpotent orbit. This
fact is important when we discuss the flavor symmetry for $A_{N-1}$ theory.

For the solution (\ref{massive}), We also need to study the moduli space of solution
of Nahm's equations with the constraints that the solution is asymptotically as (\ref{massive})
(see appendix I for more precise explanation). To understand what this space is,
we also need to apply the group theoretical result. Kronheimer also found an isomorphism between
the moduli space with the regular semi-simple orbit \cite{kro2} (see discussion in appendix I)
. In $sl(2,c)$, the regular semisimple orbit is characterized as $\det(x)=q^2$; Express
the matrix in the equation (\ref{sl2}), we find the equation $a^2+bc=q^2$, this is
the deformation of the $A_1$ space we discussed previously! From the four dimensional $N=2$
field theory point of view, the solution (\ref{massive}) corresponds to the mass-deformed theory.
 The dimension of the nilpotent orbit and the semi-simple orbit are the same, 
namely, complex dimension two; this means that a generic element in nilpotent orbit can be
deformed to a generic semisimple element.

It is time now to describe what kind of quiver tail in four dimension in which we
can find the corresponding flavor symmetry. There are two types of tail in four
dimensional $SU(2)$ quiver which give the $SU(2)$ flavor symmetry. One type
of the matter hypermultiplet is the bifundamental fields between two $SU(2)$ gauge group,
and the other type of tail is a $SU(2)$ quiver with one fundamental. They are shown in Figure 13.
\begin{figure}
\begin{center}
\includegraphics[width=3.5in,]
{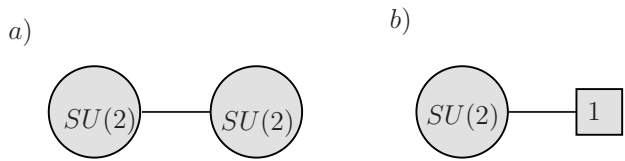}
\end{center}
\caption{ (a)Bifundamental hypermultiplet between two adjacent SU(2) gauge group gives $SU(2)$
flavor symmetry. (b) A fundamental hypermultiplet of a $SU(2)$ gauge group also gives $SU(2)$ flavor
symmetry. }
\end{figure}
For the second type of quiver tail, we may read the form from the Young tableaux associated
with the nilpotent element. The rule is that: the rank of first gauge group is the number of boxes $r_1$
in the first row, the second gauge group has the rank $r_2-r_1$, etc.
we have one single $SU(2)$ gauge group for the regular nilpotent element; we then add fundamentals to make
the theory conformal.

\subsection{Summary}

Up to now, we only consider the nilpotent element $\left(\begin{array}{cc}0&1\\
0&0\end{array}\right)$, one may ask what happened to the other nilpotent element
$\left(\begin{array}{cc}0&0\\0&0\end{array}\right)$. We can calculate the spectral curve
associated with this solution and find that there is no pole in $Tr(\Phi)^2$. We might
want to ask what is the flavor symmetry associated with this solution, as we learn from the
regular nilpotent element, we need to find the semi-simple element which has the same
dimension as the identity nilpotent element. The identity orbit is zero dimensional, and the
only semi-simple orbit with this dimension is the identity orbit itself, this means
that the identity orbit is rigid, it is against deformation to any other orbits, we
therefore claim that there is no flavor symmetry associated with this solution. According to
our rule of identifying the four dimensional quiver tail by using Young tableaux, we may associate
a quiver tail with a gauge group with rank one and another gauge group with rank 2; namely,
the quiver tail has the form $SU(2)\times SU(1)$; the $SU(1)$ gauge group might appear bizarre, but it
has brane interpretation, see discussion in \cite{SO}.  It is easy to see that there is no flavor symmetry
associated with this tail.

We only consider one singularity up to know, in general, $n$ singularities can be allowed.
$\Phi$ is a degree 1 differential on Riemann surface, so $Tr(\Phi)^2$ is a degree two
differential on Riemann surface. For massless theory, the pole structure at the singularity
is the same as local analysis. The degree $d$ differential $\Phi_d$ with the prescribed singularity has
the dimension
\begin{equation}
 \mbox{moduli of}~\phi_d=\sum_{punctures}p_d^{(i)}+3g+1-2d \label{moduli},
\end{equation}
where $(p_d^{(i)}$ is the pole structure of the $ith$ puncture, $g$ is the genus
of the riemann surface. The moduli space of the complex structure of the punctured space
is $3g+n-3$, so we have $3g+n-3$ gauge groups. Since there is one $SU(2)$ flavor symmetry for
each puncture, the theory has a total of $n$ $SU(2)$ flavor symmetry.

We can construct four dimensional $N=2$ quiver gauge theory by using the building blocks
associated with one $SU(2)$ flavor symmetry. For instance, if $g=0, n=5$, we have two gauge groups
and 5 $SU(2)$ flavor symmetries.  The quiver
gauge theory in one S-dual frame is shown in Figure 14.
\begin{figure}
\begin{center}
\includegraphics[width=3.5in,]
{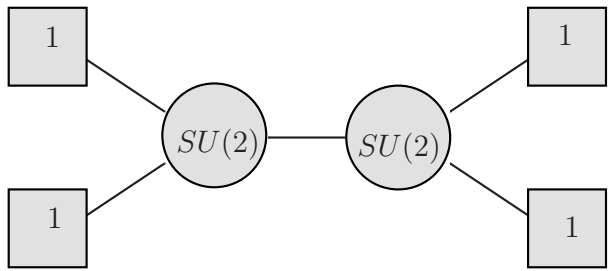}
\end{center}
\caption{One four dimensional quiver diagram with a six dimensional $A_1$ theory compactified
on a sphere with 5 punctures, we have two gauge groups and five $SU(2)$ flavor
symmetries. }
\end{figure}

The number of moduli of degree two differential has dimension $2$; The Seiberg-Witten
differential for this theory is
\begin{equation}
x^2=Tr(\Phi)^2={a_1z+a_2\over\sum_{i=1}^5(z-z_i)}.
\end{equation}
where $a_1$ and $a_2$ are identified with the dimension $two$ Coulomb branch parameters
for two gauge groups.

Here is a short summary of this section.  We have done local analysis of solution to
Hitchin's equation, and four dimensional $SU(2)$ quiver gauge theory can be
constructed from the singular solution:

1)The four dimensional theory is determined by six dimensional $(0,2)$ $A_1$ theory on the
genus g Riemann surface with punctures $n$. The number of gauge groups are the dimension of complex
structure moduli space for the punctured Riemann surface, and the number reads $3g-3+n$.

2)Hitchin's equation on Riemann surface is the dynamical equation we need to consider.
The massless theory is associated with the singular solution among which Higgs field has
only simple pole and the residue at
simple pole is a regular nilpotent element of $sl_2$ lie algebra, the mass deformed theory is associated
with Higgs field with only simple at the punctures and the residue is a semi-simple element.

3)The flavor symmetry is read from the singularity of the closure of the nilpotent orbit. For
$SU(2)$ theory, there is a $A_1$ singularity for the closure of the regular nilpotent orbit, and
the flavor symmetry is identified as $SU(2)$. The semisimple orbit is the deformation of the nilpotent
orbit.

4) The IR behavior which is encoded in Seiberg-Witten curve is described by the spectral curve
of Hitchin system. For massless theory, the spectral curve is expressed as in the form $x^2=Tr(\Phi)^2$, where
$Tr(\Phi)^2={c\over z}$ near the puncture. For massive theory, the Seiberg-Witten
curve also have the form $x^2=Tr(\Phi)^2$ but $Tr(\Phi)^2={q^2\over z^2}+{c\over z}$.

\section{$A_{N-1}$ theory}
We can generalize the $SU(2)$ results to $SU(N)$ theory. From what we
learn about $SU(2)$ theory, what's important is the singular solution to
Hitchin's equation with gauge group $SU(N)$. we conjecture that for massless
theory, the holomorphic part of the Higgs field has simple pole
at singularity and the residue is a nilpotent element of the
$sl_n$ algebra; for mass-deformed theory, the holomorphic part of
the Higgs field also has simple pole at  singularity but the
residue now is a semisimple element of $sl_n$ algebra.

\subsection{Some Mathematical Backgrounds}

We first give a short introduction to relevant mathematical results
on lie algebra structure, an readable book for physicists is \cite{nil}.
Since the pole of holomorphic part of the Higgs
field is taking value in $sl_n$, we need to consider the structure
of $sl_n$ instead of $su(n)$.

If G is a reductive group over $C$, $g$ its Lie algebra, we study the adjoint
action of G on $g$:
\begin{equation}
{\cal O}_X:=G_{ad}.X=\{\phi(X)|\phi\in G_{ad}\}.
\end{equation}
The orbits of this action
are the conjugacy classes or adjoint orbits. I

A semisimple element U of the lie algebra is an element which can
be diagonizable, an nilpotent element $U^{'}$ is an element
satisfying the relation ${U^{'}}^n=0$, where $n$ is an integer. A
conjugacy class ${\cal O}_X$ is semisimple if and only if ${\cal
O}_X={\cal O}_U$; while a conjugacy class ${\cal O}_X$ is nilpotent
if and only if ${\cal O}_X={\cal O}_{U^{'}}$.

We first define what is called a regular semisimple element in Lie
algebra. The characteristic polynomial of a matrix $X$ in $sl_n$ is
\begin{equation}
\Omega(X)=det(t-X).
\end{equation}
We can expand it as
\begin{equation}
\Omega(X)=\sum_{0\leq i\leq m}(-1)^{i}p_i(X)t^{n-i}.
\end{equation}
$p_1$ is zero since $tr X=0$. A semisimple element is called regular
semisimple if $p_l\neq 0, l\geq2$. In particular, this meas that
the diagonal elements are all different. For $sl_2$ case , we
only have the regular semi-simple orbit while for $sl_n$ case
other options are possible.

There are infinite number of semisimple
conjugacy classes and we have only finite number of nilpotent
conjugacy classes in $sl_n$ algebra. The nilpotent elements of the
$sl_n$ lie algebra are labeled by partitions of $n$ and can be put into standard
form. Introduce a partition of $n$ satisfy the conditions:
\begin{equation}
d_1\geq d_2 \geq ..\geq d_k>0~~and~~~ d_1+d_2+...+d_k=n.
\end{equation}
We label this partition as $d=[d_1,d_2,..d_k]$. We can construct
Young tableaux associated with this partition as shown in Figure 15a).
We can also construct a dual partition $d^t$ of $d$. The first row
of $d^t$ is the first column of $d$, and the second row of $d^t$ is
the second column of $d$, and so on. There is another
characterization for the dual partition: the parts of $d^t$ is given
by the following formula:
\begin{equation}
s_i=\{j|d_j\geq i\},
\end{equation}
$s_i$ equals the maximal index $j$ so that $d_j\geq i$. We also draw
a Young tableaux of the dual partition in Figure 15b).

\begin{figure}
\begin{center}
\includegraphics[width=3.5in,]
{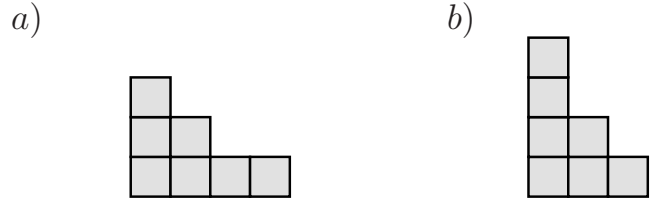}
\end{center}
\caption{ (a) Young tableaux of one partition $[4,3,1]$ of $sl_8$;
(b) The Young tableaux for transpose partition [3,2,1,1] of (a).}
\end{figure}

Each nilpotent element is labeled by a partition of $n$. It can be
put into a form using only Jordan block. The Jordan block is defined
as: given a positive integer i, we construct the $i\times i$ matrix
\begin{equation}
J_i=\left(\begin{array}{cccccc} 0&1&0&...&0&0\\
0&0&1&...&0&0\\
.&.&.&.~~~~&.&.\\
.&.&.&~.~~~&.&.\\
.&.&.&~~.~~&.&.\\
0&0&0&~...~&0&1\\
0&0&0&~...~&0&0
\end{array}\right).
\end{equation}
This matrix is called the elementary Jordan block of type $i$.

Now the nilpotent element of partition $d$ has the following form:
\begin{equation}
n=\left(\begin{array}{ccccc} J_{d_1}&0&0&...&0\\
0&J_{d_2}&0&...&0\\
.&.&.&.~~&.\\
.&.&.&~.~&.\\
.&.&.&~~.&.\\
0&0&0&...&J_{d_k}
\end{array}\right),
\end{equation}
where $J_{d_i}$ is the Jordan block with dimension $d_i$. The dimension
of this nilpotent orbits is given by
\begin{equation}
dim({\cal O}_X)=n^2-\sum_i s_i^2=n^2-1-(\sum_i s_i^2-1) \label{dimension}.
\end{equation}
Using the formula $dim({\cal O}_X)=dim(g)-dim(g^X)$,
we have $dim(g^X)=(\sum_i s_i^2-1)$ and $g^X$ is the centralizer of $X$, namely
the set of elements of lie algebra which commute with $g$. The maximal dimension occurs when the partition
is $d=[n]$, and we call it principal orbit; when $n=[2,1,1,...1]$, the
nilpotent orbit has the minimal dimension, we call it minimal orbit.

\subsection{Massless Theory and Singular Solutions to Hitchin's Equation}

After introducing those mathematical results, let's go back to
Hitchin's equation and try to find singular solutions to the equation
so that the holomorphic part of the Higgs field has simple pole at
the singularity and the residue is an $sl_n$ nilpotent element. 
In $sl_2$ case,  such a solution is found $\ref{massless}$; we can construct a similar
solution if we can find a $sl_2$ subalgebra which contains
a nilpotent element. This can be
done by establishing a homomorphism between $sl_2$
and $sl_n$ which involves a nilpotent element of the $sl_n$ algebra.

We introduce a different basis for $sl_2$ lie algebra:
\begin{equation}
H=\left(\begin{array}{cc} 1&~~0\\0&~~-1
\end{array}\right),
~~~ X=\left(\begin{array}{cc} 0&~~1\\0&~~0
\end{array}\right),
~~~ Y=\left(\begin{array}{cc} 0&~~0\\1&~~0
\end{array}\right).
\end{equation}
In this basis the nilpotent element is given by $X$,
they satisfy the commutation relation:
\begin{equation}
[H,X]=2X,~~[H,Y]=-2Y, \mbox~{and}~[X,Y]=H.
\end{equation}
For an integer $r\geq 0$, we can define a map
\begin{equation}
\rho_r:sl_2\rightarrow sl_{r+1},
\end{equation}
via
\begin{eqnarray}
\rho_r(H)=\left(\begin{array}{cccccc} r&0&0&...&0&0\\
0&r-2&0&...&0&0\\
.&.&.&.~~~~&.&.\\
.&.&.&~.~~~&.&.\\
.&.&.&~~.~~&.&.\\
0&0&0&~...~&-r+2&0\\
0&0&0&~...~&0&-r
\end{array}\right), \nonumber\\
\rho_r(X)=\left(\begin{array}{cccccc} 0&1&0&...&0&0\\
0&0&1&...&0&0\\
.&.&.&.~~~~&.&.\\
.&.&.&~.~~~&.&.\\
.&.&.&~~.~~&.&.\\
0&0&0&~...~&0&1\\
0&0&0&~...~&0&0
\end{array}\right),\nonumber\\
\rho_r(Y)=\left(\begin{array}{cccccc} 0&0&0&...&0&0\\
\mu_1&0&0&...&0&0\\
.&.&.&.~~~~&.&.\\
.&.&.&~.~~~&.&.\\
.&.&.&~~.~~&.&.\\
0&0&0&~...~&0&1\\
0&0&0&~...~&\mu_r&0
\end{array}\right),
\end{eqnarray}
where $\mu_i=i(r+1-i)$ for $1\leq i\leq r$. The homomorphism for the
nilpotent element labeled by $d$ is
\begin{equation}
\Phi_d: sl_2\rightarrow sl_n,~via~\Phi_d=\bigoplus_{1\leq i\leq k}
\rho_{d_i-1}.
\end{equation}
We can also find the commutator in $SL_n$ of this homomorphism. Assume
the nilpotent element associated with this homomorphism has the partition $d=
[d_1,d_2,...,d_k]$, let $r_i=|\{j|d_j=i\}|$, namely, $r_i$ is the number of rows with
parts $i$. The commutant is given by
\begin{equation}
G_{commu}=S(\prod_i(GL_{r_i})).
\end{equation}

Using this homomorphism, we can construct the singular solution: 
replacing $t_1,t_2,t_3$ by $\Phi_d\{t_1,
t_2,t_3\}$. The holomorphic part of the Higgs
field has a simple pole at the singularity and the residue is the
nilpotent element labeled by $d$. We want to associate these kind
of solutions with four dimensional massless $N=2$ SCFT. The first thing
we find is that we recover the result discovered
by Gaiotto: The singularity is labeled by partition of $N$ for $N=2$
$SU(n)$ SCFT. 

We next study the behavior of spectral curve near the
singularity to further confirm our conjecture. We first consider the solution
associated with the partition $[2,1,...1]$ which is the minimal orbit. We need to add the
regular term to the solution so that the holomorphic part of the
Higgs field is a regular semisimple element and looks like
\begin{equation}
\Phi(z)dz=\left(\begin{array}{cccccc}
*&({1\over z}+*)&*&...&*&*\\
{*}&*&*&...&*&*\\
.&.&.&.~~~~&.&.\\
.&.&.&~.~~~&.&.\\
.&.&.&~~.~~&.&.\\
{*}&*&*&~...~&*&*\\
{*}&*&*&~...~&*&*
\end{array}\right)dz+{\cal O}(z)dz,
\end{equation}
where $*$ is the generic numbers so that this matrix is regular
semisimple. We calculate the determinant  and expand it as a
polynomial in $x$:
\begin{equation}
\det (x-\Phi(z))=\sum_{i=2}(-1)^ip_i(z)x^{n-i}.
\end{equation}
The coefficient $p_1$ is zero since the matrix is traceless, $p_i,
i\geq2$ has simple pole at $z=0$ as we can see from calculating the
determinant. Let's recall the rule of calculating the determinant:
each term in determinant is derived by selecting numbers from the matrix,
the rule is that there is only one item selected from one row and
one column, we multiple those $n$ selected terms.

For $p_lx^{n-l}$ term in our determinant, we select $n-l$ diagonal
elements from the $\{(x-\Phi)_{22}...(x-\Phi)_{nn}\}$, we also
select ${1\over z}+*$ term from first row and a proper constant term from the second row
of $(x-\phi(z))$. We see that the coefficient $p_l$ is of order ${1\over z}$.
The result can be summarized from the corresponding
Young tableaux if we label the boxes as in Figure 16a), the pole of the
coefficient is given by $p_i=i-s_i$, where $s_i$ is the height of
the $i$th box.

\begin{figure}
\begin{center}
\includegraphics[width=3.5in,]
{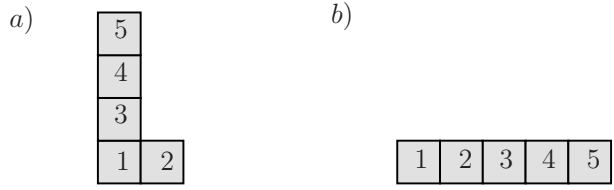}
\end{center}
\caption{ a) Young tableaux with partition $[2,1,1,1]$, the order of
poles are $p_1=1-1=0, p_2=2-1=1, p_3=3-2=1, p_4=4-3=1, p_5=5-4=1$;
b)Young tableaux with partition $[5]$, the order of poles are
$p_1=1-1=0, p_2=2-1=1, p_3=3-1=2, p_4=4-1=3, p_5=5-1=4$}.
\end{figure}

Next let's consider the solution labeled by the partition $[n]$, the
matrix $\Omega=(x-\Phi(z))dz$ (including the constant regular term)
\begin{equation}
(x-\Phi(z))dz=\left(\begin{array}{cccccc}
(x+*)&)({1\over z}+*)&*&...&*&*\\
{*}&(x+*)&({1\over z}+*)&...&*&*\\
.&.&.&.~~~~&.&.\\
.&.&.&~.~~~&.&.\\
.&.&.&~~.~~&.&.\\
{*}&*&*&~...~&(x+*)&({1\over z}+*)\\
{*}&*&*&~...~&*&(x+*)
\end{array}\right)dz+{\cal O}(z)dz.
\end{equation}
Calculate the characteristic polynomial of this matrix and leave only
the singular terms in $z$, we find that $p_i$ has pole of order $(i-1)$. To show this,
we simply expand the determinant and find the most singular term for the coefficient.
For term $p_ix^{n-i}$,  we
select $(i-1)$ $({1\over z}+*)$ terms just above the diagonal terms, and then
select the remaining $n-i$ diagonal terms, this is the maximal pole we can
get at $z=0$. The order of pole can be read from the Young tableaux, namely
$p_i=i-s_i$, see Figure 16b).

For general partition, the matrix $(x-\Phi(z))$ has the form:
\begin{equation}
(x-\Phi(z))dz=\left(\begin{array}{ccccc} I_{d_1}&*&*&...&*\\
{*}&I_{d_2}&*&...&*\\
.&.&.&.~~&.\\
.&.&.&~.~&.\\
.&.&.&~~.&.\\
{*}&*&*&...&I_{d_k}
\end{array}\right)dz+{\cal O}(z)dz.
\end{equation}
where $I_{d_i}$ takes the form
\begin{equation}
I_{d_i}=\left(\begin{array}{ccccc} x+*&({1\over z}+*)&*&...&*\\
{*}&x+*&({1\over z}+*)&...&*\\
.&.&.&.~~&.\\
.&.&.&~.~&.\\
.&.&.&~~.&({1\over z}+*)\\
{*}&*&*&...&x+*
\end{array}\right)
\end{equation}
The orders of pole for the coefficients $p_i,2\leq i\leq d_1$  are calculated as follows:
we choose the diagonal terms from the other blocks except the first block $I_{d_1}$,
then we do the same analysis on the first block as we do on the partition $[n]$;
the order of pole is given by $i-1$ when
$i\leq d_1$. To calculate term $p_{d_1+1}x^{n-d_1-1}$,
We select $d_1-1$ terms of form ${1\over z}$ and a constant term from first
block $I_{d_1}$; We can not
choose another ${1\over z}$ term, since if we choose a ${1\over z}$ term say coming from the first row
from the second block, we can not choose the
two diagonal terms adjacent to it in calculating the determinant, the maximal order of $x$ we can get is $n-d_1-2$.
Therefore, the order of pole is $d_1-1$, or $d_1+1-2$. The order of poles for
other terms $p_i, d_1<I\leq d_2$ is given by $i-2$. We do the same analysis when we jump
from $d_i$ to $d_{i+1}$, in general, the order of pole is read from the Young tableaux and
given by $i-s_i$, where $s_i$ is the height of the $ith$ box. This exactly matches
the result of Gaiotto \cite{Gaiotto1}, see the discussion in section 2.

\subsection{Flavor symmetry}
Next, we want to analyze the flavor symmetry associated with singularity with nilpotent residue. We first
analyze the principal nilpotent orbit. According to our previous discussion on $A_1$ theory,
we can understand the flavor symmetry from the singularity of the moduli space of solution
with the prescribed boundary behavior. There is an isomorphism between the moduli space of solutions
and nilpotent orbit. In the present case, We need to study the geometry of the principal nilpotent orbit
or more precisely we need to consider the closure of this orbit. Let's denote this orbit as
$C_{reg}$, and its closure $\bar{C}_{reg}$ is the set of all nilpotent elements. The boundary
$\partial C_{reg}:=\bar{C}_{reg}-C_{reg}$ is the closure of the conjugacy class $C_{(n-1,1)}$,
where $d=[n-1,1]$ is the partition of this nilpotent class. This class is called the subregular
nilpotent conjugacy class and will be denoted by $C_{subreg}$. From our formula counting the dimension
(\ref{dimension}), we have $codim_{\bar{c}_{reg}}C_{subreg}=2$. It might be useful for our understanding
if we recall in $sl_2$ case, the principal nilpotent orbit is $O_{Y_1}$, and its closure contains
$O_{Y_2}$ which is the subregular nilpotent class and has codimension two in the closure of $O_{Y_1}$.

The following result is due to Brieskorn\cite{brie}: The singularity of $\bar{C}_{reg}$ in $C_{subreg}$ is
smoothly equivalent to the simple surface singularity $A_{n-1}$:
\begin{equation}
Sing(\bar{C}_{reg}, C_{subreg})=A_{n-1}.
\end{equation}
As usual $A_{n-1}$ denotes the isolated singularity given by the equation $x^n+y^2+z^2=0$. So we
can identify the flavor symmetry associated with this solution as $SU(n)$.

Next, let's consider the flavor symmetry associated with the minimal nilpotent element.
Let's denote the conjugacy class of this element as $C_{min}=C_{(2,1,....1)}$, it has dimension
 $dim~C_{min}=2(n-1)$. We have $\bar{C}_{min}=C_{min}\bigcup\{0\}$, where $\{0\}$
is the identity orbit. We have the following result about the closure: $\bar{C}_{min}$ has a isolated
rational singularity in zero.

To understand this singularity, we can describe a resolution of the singularity in $\bar{C}_{min}$.
Let $P\subset GL_n$ be the stabilizer of the line $e_1$, $e_1:=(1,0,...,0)$, and denote
by $n$ the nilradical of the parabolic subalgebra $LieP$ of the all the nilpotent orbits.
Then $GL_n/P\cong P^{n-1}$ and the associated vector bundle
\begin{equation}
GL_n(k)\times n\rightarrow GL_n/P,
\end{equation}
is the cotangent bundle. Furthermore $n\subset\bar{C}_{min}$ and the canonical map
\begin{equation}
\phi:GL_n\times n\rightarrow\bar{C}_{min},
\end{equation}
induced by $(g,A)\rightarrow gAg^{-1}$ is a resolution of singularities (i.e. is proper and birational)
with $\phi^{-1}(0)=$ zero section of the cotangent bundle. This means that we obtain the singularity
$\bar{C}_{min}$ by "collapsing" the cotangent bundle of $P^{n-1}$. We call this singularity by
$a_{n-1}$:
\begin{equation}
Sing(\bar{C}_{min},0)=a_{n-1}.
\end{equation}
We claim that the flavor symmetry associated with this singularity is $U(1)$ with only one exception
$a_1$. Since $a_1=A_1$, in that case, the flavor symmetry is enhanced to $SU(2)$.

To understand the flavor symmetry associated to the general partition $d=[d_1...d_k]$, we can
go along the same line as the principal orbit and minimal orbit
 with some complication.  We follow \cite{sing} to illustrate the main point. First,
we define an order relation among the nilpotent orbits. Given
two partitions $\eta=(p_1,p_2,...p_s)$ and $\nu=(q_1,q_2....q_t)$ of $n$, we say $\eta\geq \nu$ if

\begin{equation}
\sum_{i=1}^jp_i\geq\sum_{i=1}^jq_i~~\mbox{for all j}.
\end{equation}

If $\eta>\nu$ and no partition is in between them (i.e. $\eta$ and $\nu$ are adjacent
in the ordering), then the Young tableaux of $\nu$ is obtained from $\eta$ by moving one
box up either to the next row or to the next column, see Figure 17 for the illustration.
\begin{figure}
\begin{center}
\includegraphics[width=3.5in,]
{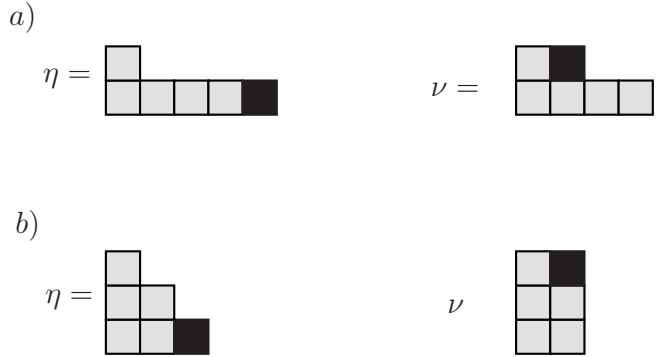}
\end{center}
\caption{ a)$\nu$ is derived from $\eta$ by moving a box up to the next row;
b) $\nu$ is derived from $\eta$ by moving a box up to the next column}.
\end{figure}
There might be more than one adjacent partition to $\eta$, see an example in Figure 18.
\begin{figure}
\begin{center}
\includegraphics[width=3.5in,]
{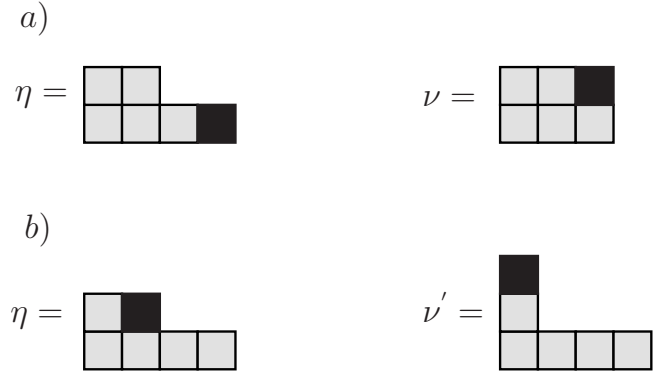}
\end{center}
\caption{We can find different adjacent Young diagram by moving different boxes.
 a)$\nu$ is derived from $\eta$ by moving a box up to the next row;
b) $\nu$ is derived from $\eta$ by moving a different box up to the next row.}
\end{figure}
Formally, the adjacent pair $\eta=(p_1,p_2,....,p_s)$ and $\nu=(q_1, q_2,...,q_t)$
can be expressed as the following two types:

I) If we need to move up a box to next row of $\eta$, then there is a an integer $i$,
such that $q_k=p_k$ for $k\neq i,i+1$ and $q_i=p_i-1\geq q_{i+1}=p_{i+1}+1$.

II) If we need to move up a box to next column of $\eta$, then there are integers $i,j$
such that $p_k=q_k$ for $k\neq i,j$ and $q_i=p _i-1=q_j=p_j+1$.

Given two partitions $\eta$ and $\nu$ of $n$, $\eta\geq\nu$ if
and only if $\bar{C}_\eta\supseteq C_\nu$. We call a degeneration $C_\nu\subseteq
\bar{C}_\eta$ minimal if $C_{\nu}$ is open in $\bar{C}_\eta-C_\eta$, i.e. if $C_\nu\neq C_\eta$
and there is no conjugacy class $C$ such that $\bar{C}_{\nu}\subseteq\bar{C}\subseteq\bar{C}_\eta$.
This means that $\eta$ and $\nu$ are adjacent.

Let $C_\nu\subset\bar{C}_\eta$ be a minimal degeneration, we have one of the following two
cases:

I) $codim_{\bar{C}_\eta}C_{\nu}=2$ and the Young tableaux of $\nu$ is obtained from
$\eta$ by moving one box up to the next row.

II) $codim_{\bar{C}_\eta} C_{\nu}=2r$ and the Young tableaux of $\nu$ is obtained
from $\eta$ by moving one box up to the next column. If the box is moved from the
i-th row to the j-th row, then $r=j-i$.

We can find the singularity of $\bar{C}_\eta$ its minimal degeneration $C_\nu$
by exploring the reduction relation. Let $C_\nu\subseteq\bar{C}_\eta$ be a
degeneration of nilpotent conjugacy classes and assume the first $r$ rows and the first
columns of $\eta$ and $\nu$ coincide. Denote by $\eta^{'}$ and $\nu^{'}$
the Young diagrams obtained from $\eta$ and $\nu$ by erasing these rows and
columns, then $C_{\nu^{'}}\subset\bar{C}_{\eta^{'}}$,
\begin{equation}
codim_{\bar{C}_\eta^{'}}C_{\nu^{'}}=codim_{\bar{C}_\eta}C_{\nu}~~~and~~Sing(\bar{C}_{\eta^{'}},
C_{\nu^{'}})=Sing(\bar{C}_\eta, C_\nu)
\end{equation}

Using the reduction formula, We can determine the singularity for closure of
any nilpotent orbit on a codimension two orbit which lies in the closure. Let $C^{'}\subseteq\bar{C}$
be a minimal degeneration of nilpotent conjugacy classes, the singularity of $
\bar{C}$ in $C^{'}$ is either of type $A_m$ or is of type $a_l$. More accurately,
\begin{eqnarray}
Sing(\bar{C},C^{'})=A_{m-1}~~for~~some~~m<n~~if~~codim_{\bar{C}}C^{'}=2 \nonumber\\
Sing(\bar{C},C^{'})=a_{l}~~if~~codim_{\bar{C}}C^{'}=2l\geq2
\end{eqnarray}

We give a heuristic proof of above theorem and in the process of the proof we will
see how to determine $m$ and $l$. Let $\eta$ and $\nu$ be the associated partitions
to $C$ and $C^{'}$, $\eta=(p_1,p_2,..,p_s)$. If $\nu$ is derived from $\eta$ by moving
up a box to next row, then $\nu=(p_1,..p_i-1,p_{i+1}+1,..p_s)$ for some $i$.  $\eta$
and $\nu$ have the same $i-1$ rows and the same first $p_{i+1}$ columns by examining their partitions; After erasing
these same rows and columns, the Young diagram of  $\eta^{'}$ has the total boxes $m=p_i-p_{i+1}$,
and it is of the type $\eta^{'}=[m]$, and the Young diagram of $\nu^{'}$ also
has total boxes $m$ and it is of the type $\nu^{'}=[m-1,1]$,  the following relation holds
\begin{equation}
Sing(\bar{C}, C^{'})=Sing(\bar{C}_{(m)},C_{(m-1,1)})=A_{m-1}
\end{equation}
One example of this type is given in Figure 19a).

If $\nu$ is derived from $\eta$ by moving a box to next column, then $\nu=(p_1,...p_i-1,p_{i+1},...
p_j+1,p_{j+1},...p_s)$, for some $i<j$ with $p_i-1=p_{i+1}=...=p_j+1$. The first $i-1$
rows of  $\eta$ and $\nu$ are the same, and the first $p_j$ columns of them are the same. After
erasing these same rows and columns, $\eta^{'}$ has total boxes $j-i+1$ and has the partition
$\eta^{'}=[2,1,1,..1]$, and $\nu^{'}$ has the same total of boxes and has the partition
$\nu^{'}=[1,1,1,...1]$. Now the singularity is determined as
\begin{equation}
Sing(\bar{C}, C^{'})=Sing(\bar{C}_{(2,1,..,1)},0)=a_{j-i}
\end{equation}
and $codim_{\bar{C}}C^{'}=2(j-i)$. One example of this type is given in Figure 19b).
\begin{figure}
\begin{center}
\includegraphics[width=3.5in,]
{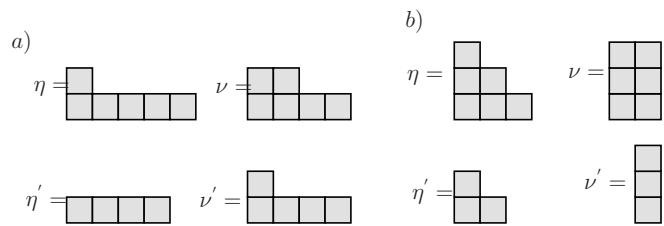}
\end{center}
\caption{ a)$\nu$ is derived from $\eta$ by moving a box up to the next row, after
erasing the first same set of rows and the first same set of columns, we find the singularity is
$A_4$.
b) $\nu$ is derived from $\eta$ by moving a different box up to the next row, we find
the singularity is $a_2$}.
\end{figure}

In fact, the minimal singularities of can be read from  Young tableaux
of the orbit itself. The rule is stated in the above proof process. As we noted before, it is possible for
a nilpotent orbit to have more than one minimal degeneration. We give a table  of $sl_6$
which illustrates the singularities for the minimal degeneration in Figure 20.
\begin{figure}
\begin{center}
\includegraphics[width=3.5in, height=3in]
{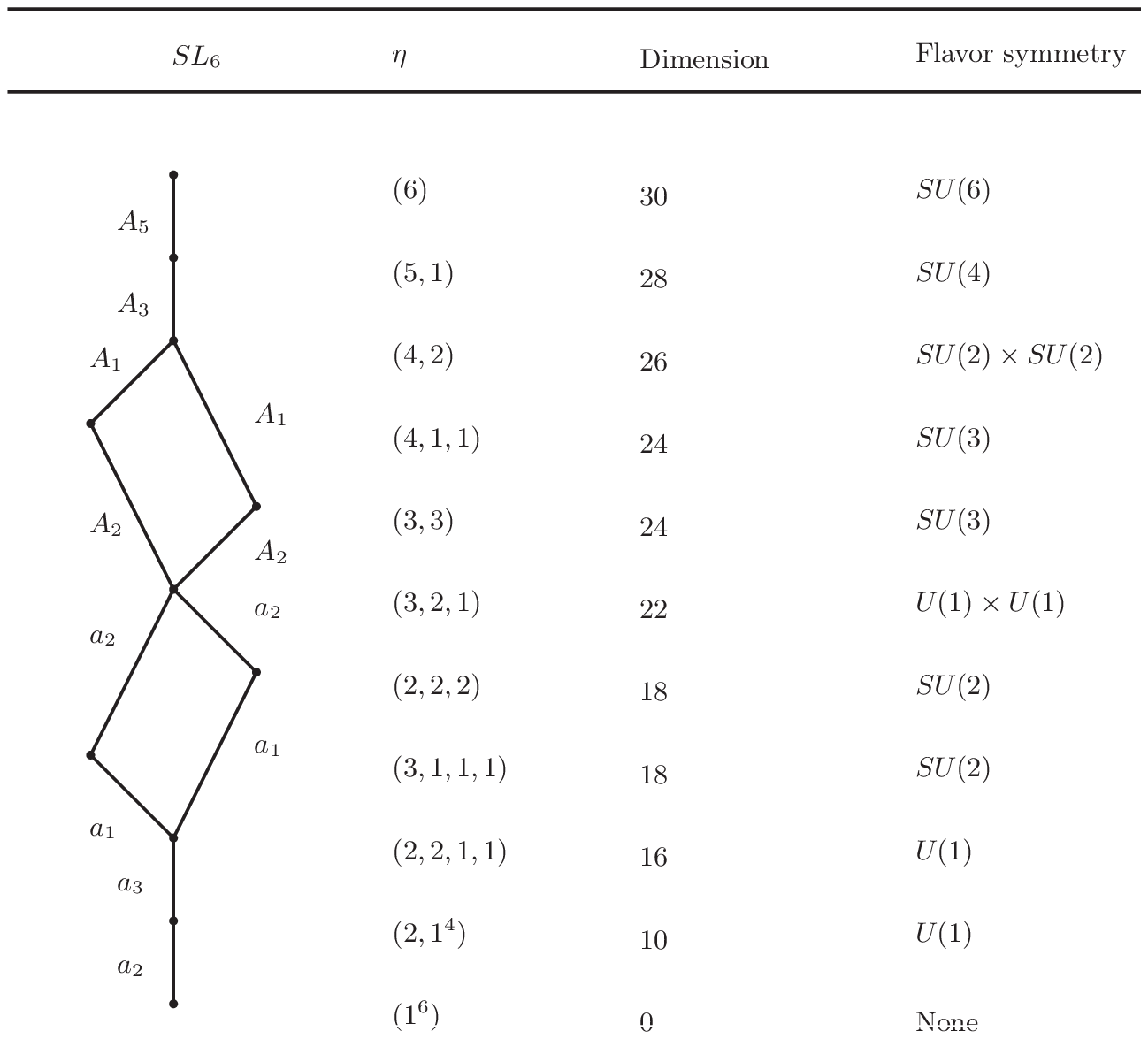}
\end{center}
\caption{The singularity of closure of nilpotent orbit on its minimal degeneration of $sl_6$ lie algebra. }
\end{figure}

Now we can state what kind of flavor symmetries are associated with a singular solution of  Hitchin's
equation with nilpotent residue:

The flavor symmetries associated with a nilpotent orbit are determined by the singularity of its closure on
its minimal degeneration, i.e, if the singularity is of type $A_{m-1}$, we have $SU(m)$ flavor symmetry;
If the singularity is of type $a_m, m>1$, the flavor symmetry is $U(1)$; If the singularity is $a_1$,
the flavor symmetry is enhanced to $SU(2)$.  If there are more than one degeneration, the flavor symmetry
are from the product of all the singularities.

Let's state some examples from $sl_6$ lie algebra to illustrate the idea. We can read the flavor symmetries
from Figure 20 for all the nilpotent elements. To get the real flavor symmetry for the four dimensional 
theory, there are subtleties about $U(1)$ factors. Since we study the singularity of the three dimensional
coulomb branch, the information about $U(1)$ flavor symmetry of four dimensional theory is usually lost. 
For the partition $[n]$ and $[2,1,...1]$, there is no ambiguity, the flavor symmetry is $SU(n)$ and $U(1)$
respectively. For other generic partition, if the singularity type is from the degeneration by moving the box from
the highest row, we do not add a $U(1)$ factor, otherwise, we need to include a $U(1)$ factor. This rule is also
true for partition $[n]$. One can check that this 
really recovers the flavor symmetry described by Gaiotto. The reason why the $U(1)$ symmetry is not included in this case is that for the fundamentals attached on the gauge group $SU(N)$, the $U(1)$ factor is denoted by the simple
puncture.

It is the same as in $SU(2)$ case to determine what kind of basic building block
we can derive for a given singularity. If the nilpotent element associated with
the singularity has the partition $d=[d_1,d_2,...d_k]$, the tail has gauge group $SU(d_1)\times SU(d_2+d_1)
..\times SU(N)$, and we add fundamental hypermultiplets to gauge group
to make the theory conformal. When we have the minimal orbit, namely the partition $d=[2,1,...1]$,
the bifundamentals can be also represented by this orbit.
\subsection{Mass Deformed Theory}
In this subsection, we are going to study what kind of singular solutions correspond to mass-deformed
theory. Let's recall what we learned about $SU(2)$ theory. The massless theory is
associated with  Higgs field whose residue is a nilpotent element $Y_1$
labeled by the partition $d=[2]$; what is actually important
is the moduli space of solutions with appropriate boundary conditions so that
the residue is living in the conjugacy class of $Y_1$. There is an isomorphism between
this moduli space and nilpotent orbit itself. On the other hand, the mass deformed theory
is described by a solution to Hitchin's equation so that the residue of the Higgs field
is a semisimple element (which is also regular for $su(2)$). We also are concerned about
the moduli space of solutions and there is also an isomorphism between the space of solutions
and the semi-simple orbit itself. Both nilpotent orbit and semi-simple orbit
are hyper-Kahler manifolds and the closure of nilpotent orbit is singular and
semi-simple orbit can be thought of the deformation of nilpotent orbit. The basic
requirement for this understanding is that they must have the same complex dimensions.

Generalizing  above considerations of $SU(2)$ to $SU(N)$, we
 need to find certain kinds of solutions of Nahm's equation
  whose moduli space is a deformation of the moduli space of solutions we are studying in the last subsection.
In general, given a triple $(\tau_1,\tau_2,\tau_3)$, let $\sigma_1,\sigma_2,\sigma_3$ be
elements of $g$ which commute with $\tau_j$ and which satisfy the $su(2)$ relations,
a solution to the equation is
\begin{equation}
a=\tau_1+{\sigma_1\over 2s},~~b=\tau_2+{\sigma_2\over 2s},~~a=\tau_1+{\sigma_3\over 2s},~~s\rightarrow\infty.
\end{equation}
These conditions means that the residue of the Higgs field takes value in $\tau_2+i\tau_3+\sigma^c$, where
$\sigma^c$ is the nilpotent element we can get from $su(2)$ algebra $\sigma_1,\sigma_2,\sigma_3$.
There is a one-to-one correspondence between the solution space with this boundary conditions and
the adjoint orbit which contains $\tau_2+i\tau_3+\sigma^c$ (see appendix I for more details), it is
also proved that this space is a hyper-Kahler manifold.

Since the nilpotent orbit is identified with the massless theory, so we are led to think
that the mass-deformed theory corresponds to semisimple orbit. The question is to identify the semi-simple
orbit, we will call those semi-simple orbits as the mass-deformed orbits.
The closure of the nilpotent orbit is singular and we can think of the mass-deformed
orbit as the deformation of the closure. The necessary condition for this is that the mass-deformed
orbit has the same dimension as the closure of the nilpotent orbit.

The dimension for a nilpotent orbit is given by (\ref{dimension}). The following lemma can be used to calculate dimension of a semi-simple orbit:

Let $g$ be a reductive lie algebra and $X$ is element in a semisimple orbit, its
centralizer $g^X$ is reductive and there exists a Cartan subalgebra $h$ containing $X$.
If $\Phi$ denotes the roots for the pair $(g,h)$, then $g^X=h\bigoplus\sum_{\alpha\in\Phi_X}
g_\alpha$, where $\phi_X=\{\alpha\in\phi|\alpha(X)=0\}$.

We study $sl_3$ as an example to show how to use the above lemma to calculate the dimension
of a semisimple orbit. The traceless diagonal matrices in $sl_3$, denoted as $h$, form a three dimensional
Cartan subalgebra. For each $1\leq i \leq 3$, define a linear functional in the dual
space $h^{*}$ by
\begin{equation}
e_i\left(\begin{array}{clr}
h_1&0&0\\
0&h_2&0\\
0&0&h_3\\
 \end{array}\right)=h_i.
\end{equation}
The standard choices of positive and simple roots are
\begin{equation}
\Phi^+=\{e_i-e_j|1\leq i<j\leq3\}~and~\triangle=\{e_i-e_{i+1}|1\leq i\leq 2\}.
\end{equation}
Consider the following matrices
\begin{equation}
X_1=\left(\begin{array}{ccc}
m_1&0&0\\
0&m_2&0\\
0&0&-(m_1+m_2)\end{array}\right),~~
X_2=\left(\begin{array}{ccc}
m_1&0&0\\
0&m_1&0\\
0&0&-2m_1)\end{array}\right),~~
X_3=\left(\begin{array}{ccc}
0&0&0\\
0&0&0\\
0&0&0\end{array}\right)~~.
\end{equation}
We now describe how to calculate dimension of semi-simple orbit ${\cal O}_{X_k}$ for
$1\leq k\leq 3$.

For case $X_1$, since $\alpha(X_1)\neq0$ for any simple
roots $\alpha\in \triangle$,  $g^{X^1}$ is  a Cartan subalgebra using our lemma.
The dimension for $X_1$ is
\begin{equation}
dim({\cal O}_{X_1})=dim(g)-dim(g^{X^1})=8-2=6.
\end{equation}

For case $X^2$,  $\alpha(X_2)=0$ if and only if $\alpha=\pm(e_1-e_2)$, so
$g^{X_2}=h\bigoplus g_{e_1-e_2}\bigoplus g_{e_2-e_1}$ and $dim({\cal O}_{X_2})=8-4=4$.

For case $X^3$, $\Phi_{X_3}=\{\pm(e_1-e_2),\pm(e_2-e_3),\pm(e_1-e_3)\}$, then $dim(g^{X_3})=8$,
and $dim({\cal O}_{X_3})=0$.

Let's study the semi-simple orbits for general $sl_n$ algebra. We want to study
semi-simple elements labeled by a partition $d=[d_1,d_2,...d_k]$ of n. It has
the form $X_{d}=diag(m_1,..m_1,m_2...m_2,....m_k..m_k)$, where the first $d_1$ diagonal terms
have the same value, etc. It is interesting that we can also label semi-simple orbits by
partitions of n. The dimension of the orbit ${\cal O}_{X_{d}}$ can be calculated
by using the lemma we introduced above.

Let $h$ be traceless diagonal $n\times n$ matrices; Define the linear functional $e_i\in h^{*}$
by $e_i(H)=i^{th}$ diagonal entry of H, here $ 1\leq i\leq n$. The root system is
$\{e_i-e_j|1\leq i,j\leq n, i\neq j\}$ in this representation. Elements in $\phi(X_d)$ from first block are
\begin{equation}
\Phi^1(X_{d})=\{\pm(e_1-e_2),\pm(e_1-e_3),.\pm(e_1-e_{d_1}), \pm(e_2-e_3) \pm(e_2-e_4)..\pm(e_2-e_{d_1})...
\pm(e_{d_1-1}-e_{d_1})\}
\end{equation}
For the other block we can similarly find the other roots which satisfy the condition $\alpha(X_d)=0$

The dimension of the centralizer of $X_{d}$ is
\begin{equation}
dim(g^{X_{d}})=n-1+2(d_1-1+d_1-2+...+1)+2(d_2-1+d_2-2+..+1)+...+
(2d_k-1+...+1),
\end{equation}
sum them up, we have
\begin{equation}
dim(g^{X_{d}})=n-1+
\sum_i^kd_i(d_i-1)=n-1+\sum_i^kd_i^2-n=\sum_i^k d_i^2-1.
\end{equation}
where the condition $\sum_i^kd_i=n$ is used. The dimension of the semisimple orbit
is
\begin{equation}
dim({\cal O}_{X_d})=n^2-1-\sum_i^k d_i^2+1=n^2-\sum_i^k d_i^2.
\end{equation}

Recall the dimension (\ref{dimension}) of a nilpotent orbit with the partition $d_1$:
\begin{equation}
dim({\cal O}_{X_{d_1}})=n^2-\sum_i s_i^2.
\end{equation}
Where $s^i$ is the rows of the dual partition of $d_1$.

Comparing the dimension of the nilpotent orbits and dimension of the semi-simple
orbits, we have the following observation of the mass-deformed theory for a puncture labeled 
 by the partition $d$:

The mass deformed theory is described by the singular solution of Hitchin's equation;
The Higgs field has simple pole at the singularity and whose residue is a semisimple
element with the form labeled by the dual partition $d^t=[d^t_1,d^t_2,...d^t_k]$ of $d$:
\begin{equation}
\Phi(z)dz={dz\over z}diag(m_1,..m_1,m_2,..m_2,....m_k...m_k)+...,
\end{equation}
where $\Phi$ has  $d_1$ $m_1$ eigenvalues, $d_2$ $m_2$ eigenvalues and so on. The Seiberg-
Witten curve is the spectral curve of the Hitchin's system.

We can read flavor symmetry from the dual partition $d^t$. For this
partition, we can also define a $sl_2$ homomorphism 
and the commutant of this homomorphism
in $sl_n$ is given by

\begin{equation}
G_{commu}=S(\prod_i(GL_{r_i}),
\end{equation}
where $r_i$ is the number of rows of $d_t$ with boxes i. This number $r_i$ is also
the number of columns of $d$ with heights i. Using its real form, we identify this as
the flavor symmetry associated with the puncture and this agrees with Gaiotto's result.
See \cite{SO} for the relevant discussion.

\section{Conclusion}
In this paper, we argue that it is the Hitchin's equation which determines not only IR limit
but also the UV theory of the four dimensional
$N=2$ superconformal field theories. We study the local singular solution of Hitchin's equation so
that the Higgs field has a simple pole at the singularity. We show that the massless theory is associated
with the solution so that Higgs field has nilpotent residue and the mass deformed
theory is associated with the semi-simple residue; The moduli space of solutions
of the mass deformed theory can be thought of as the deformation of the moduli space
of the massless theory. The Seiberg-Witten curve which determines the IR behavior is
given by the spectral curve of Hitchin's equation.

It is interesting to extend our analysis to six dimensional $D_N$ theory compactified
on a Riemann surface. The four dimensional theory can be derived by adding O4 planes to
the brane configurations we considered in this paper \cite{SO1,SO2}.
The flavor symmetry and the Young tableaux classification is given
by Tachikawa \cite{SO}. In that paper, the author shows that there are
 two different kinds of tails which are related to $USp$
and $SO$ group.  We might need to study  Hitchin's equation with $SO(2n)$ group, we
have almost the same group structure associated with the nilpotent orbits and semisimple orbits.
However, it is puzzling why we need $USp$ and $SO$ groups at different type of singularities.

Similar analysis can be done on six dimensional $E_N$ theory on punctured Riemann surface.
We don't have any brane configuration so we don't have any four dimensional understanding
about the field theory. However, following the same analysis from this paper, we hope we
can learn something about this type of theories.

In this paper, we only carry the local analysis of solutions to Hitchin's equation,
it is interesting to extend the analysis to consider the global constraint.
we also only study the solutions with simple poles at the singularity. We
can also studied the solutions with higher order singularities. This type of solutions
can be also used to study the SCFT and it can be used to represent the asymptotically free
theory. We leave this for the future study.

Nahm's equation plays a fundamental role in our study. It is also used in an essential
way to study boundary conditions for $N=4$ theory by Gaiotto and Witten \cite{GW1, GW2}.
In those papers, Nahm's equation is important to get three dimensional mirror symmetry \cite{KS}.
Since the moduli space of Hitchin's equation is closed related to 
coulomb branch of three dimensional theory, we may wonder whether we can find new three
dimensional mirror pair by compactifying four dimensional $N=2$ theory down to three dimension. This
is indeed the case \cite{new}.

Finally, Hitchin's equation is related to the KZ(Knizhnik-Zamolodchikov) equation of the two dimensional conformal
field theory, this shed the light about why $AGT$ \cite{Gaiotto2} relation is possible, we hope we can
learn more about the relation between four dimensional SCFT and two dimensional CFT by
studying Hitchin's equation.

\begin{flushleft}
\textbf{Acknowledgments}
\end{flushleft}
This research was supported in part by the Mitchell-Heep chair in
High Energy Physics (CMC) and by the DOE grant DEFG03-95-Er-40917.

\begin{flushleft}
\textbf{Appendix I}
\end{flushleft}
The Hitchin's equation defined on a Riemann sruface is
\begin{eqnarray}
F-\phi\wedge\phi=0\nonumber\\
d_A\phi=0,~
d_A*\phi=0,
\end{eqnarray}
where $d_A$ is the covariant derivative and $*$ is the Hodge star operator. Define
the local coordinates $z=re^i\theta$, We
are looking for the rotational invariant solution with the form:
\begin{eqnarray}
A=a(r)d\theta+f(r){dr\over r}\nonumber\\
\phi=b(r){dr\over r}-c(r)d\theta,
\end{eqnarray}
with the functions $a,b,c,f$ which take values in lie algebra $g$ of $G$. Though $f$ can
be gauged away, we keep it for later use. The space of solutions with appropriate
boundary conditions has a hyper-Kahler structure. One way to understand this fact is
to think of the functions $(f,a,b,c)$ as giving a map from the open unit interval
to the quarternions $H\cong R^4$, tensored with $g$. The hyper-Kahler structure
comes from the hyper-Kahler structure on $H$. In one complex structure, $b+ic$ is
the complex structure parameters and $f-i\alpha$ is the kahler structure parameters.
The others can be obtained by applying an $SO(3)$ rotation to the triple $(a,b,c)$.

If we consider a local region around the singularity $r\in [0,1]$, and we
define another coordinate $s=-\ln r$ and in this new coordinate the range
is $s\in [0,\infty)$. Define $D/Ds=d/ds+[f,.]$, Hitchin's equation becomes
\begin{eqnarray}
{Da\over Ds}=[b,c]\nonumber\\
{Db\over Ds}=[c,a]\nonumber\\
{Dc\over Ds}=[a,b],
\end{eqnarray}
when $f=0$, the above equation becomes the Nahm's equation. The above equations
is invariant under gauge transformations:
\begin{eqnarray}
f\rightarrow Ad(g)(f)-{dg\over ds}g^{-1}\nonumber\\
a\rightarrow Ad(g)a,~~b\rightarrow Ad(g)b,~~c\rightarrow Ad(g)c.
\end{eqnarray}
where $g:[0,\infty)\rightarrow G$ is any path. Introducing the complex variables
\begin{equation}
\alpha(s)={1\over 2}(f(s)+ia(s)),~~~~\beta(s)={1\over 2}(b(s)+ic(s)),
\end{equation}
the Nahm's equation can be rewritten as one 'real' equation and one 'complex'
equation:
\begin{equation}
{d\over ds}(\alpha+\alpha^*)+2([\alpha,\alpha^*]+[\beta,\beta^*]=0 \label{real},
\end{equation}
\begin{equation}
{d\beta\over ds}+2[\alpha,\beta]=0 \label{complex}.
\end{equation}

\begin{flushleft}
\textbf{Boundary Conditions, Isomorphism between Moduli Space of Solutions
and Nilpotent Orbits}
\end{flushleft}
We are studying the Nahm's equation with appropriate boundary conditions. Let
$\rho:su(2)\rightarrow g$ be a Lie algebra homomorphism and write
\begin{equation}
H=\rho\left(\begin{array}{cc}
1&0\\
0&-1\end{array}\right),~~~
X=\rho\left(\begin{array}{cc}
0&1\\
0&0\end{array}\right),~~~
Y=\rho\left(\begin{array}{cc}
0&0\\
1&0\end{array}\right).
\end{equation}

Let's denote $M(\rho)$ as the space of smooth solutions $\alpha (s),\beta (s):[0,\infty)\rightarrow g^c$ which
satisfy the following boundary conditions:

i)\begin{equation}
2s\alpha(s)\rightarrow Ad(g)H,~~~s\beta(s)\rightarrow Ad(g_0) Y,~~s\rightarrow \infty,
\end{equation}
with some $g_0\in g$.

ii)\begin{equation}
2s\alpha(s)\rightarrow 0,~~s\beta(s)\rightarrow 0,~~s\rightarrow 0.
\end{equation}
It is proven by Kronheimer \cite{kro1} that $M(\rho)$ has a hyper-kahler structure and is
isomorphic to the nilpotent orbit associated with $Y$.

\begin{flushleft}
\textbf{Isomorphism between Moduli Space of Solutions and Semisimple orbits}
\end{flushleft}

Let's first define a triple $(\tau_1,\tau_2,\tau_3)$ which lives in the Cartan subalgebra
( $(\tau_1,\tau_2,\tau_3)$ are denoted as $(\alpha,\beta,\gamma)$ in our main discussion).
We first assume that the triple is regular in the sense that the union of the centralizer
$C(\tau_1)\bigcup C(\tau_2) \bigcup C(\tau_3)$ is a Cartan subalgebra.

let $M(\tau_1,\tau_2,\tau_3)$ be the space of solutions $(a(s), b(s), c(s))$, satisfying the
boundary condition
\begin{equation}
lim_{s\rightarrow\infty}a(s)=Ad(g_0)\tau_1,~~lim_{s\rightarrow\infty}b(s)=Ad(g_0)\tau_2,
~~lim_{s\rightarrow\infty}c(s)=Ad(g_0)\tau_3.
\end{equation}
for some $g_0\in g$. This space has a hyper-kahler structure.

If we further require $(\tau_2,\tau_3)$ be a regular pair, and suppose that $(a(s),b(s),c(s))$
is a solution of Nahm's equation (we set $f=0$ at this step)
 belonging to $M(\tau_1,\tau_2,\tau_3)$. If we set
$\sigma_1=a(0),\sigma_2=b(0),\sigma_3=c(0)$, we have the following statement:

\emph{The element $\sigma_2+i\sigma_3\in g^c$ belongs
to the same complex adjoint orbit as $\tau_2+ i\tau_3$. }

We can prove above statement by using equation (\ref{complex}). Consider the solution $(\alpha, \beta)$
of that equation which is obtained from $(a(s),b(s),c(s))$ by setting $f(s)=0$, 
the equation implies that the path $2\beta(s)$ lies entirely within a single adjoint orbit
of the complex group. let ${\cal O}(2\beta)$ denote this orbit, and let ${\cal O}(\sigma)$ and
${\cal O}(\tau)$ be the orbits which contain $(\sigma_2+i\sigma_3)$ and $\tau_2+i\tau_3$ respectively.
The boundary conditions are
\begin{eqnarray}
2\beta(s)\rightarrow \tau_2+i\tau_3,~~as~s\rightarrow \infty \nonumber\\
2\beta(s)\rightarrow \sigma_2+i\sigma_3,~~as~s\rightarrow 0,
\end{eqnarray}
so the closure of ${\cal O}(2\beta)$ contain ${\cal O}(\sigma)$ and ${\cal O}(\tau)$, but
${\cal O}(\tau)$ is a regular semisimple orbit, which means that it is closed and is
contained in the closure of no other orbit. It follows that ${\cal O}(2\beta)={\cal O}(\sigma)
={\cal O}(\tau)$.

From this point, we denote ${\cal O}$ the adjoint orbit which contains
$\tau_2+i\tau_3$. The above statement provides us with a map
\begin{equation}
f:M(\tau_1,\tau_2,\tau_3)\rightarrow {\cal O}~~~(a(s),b(s),c(s))\rightarrow (b(0)+c(0)),
\end{equation}
Kronheimer has proved that this is a bijection \cite{kro2}.

For a non-regular triple, the boundary conditions can be different. Let $\tau_1,\tau_2,\tau_3$
be non-regular elements of the Cartan algebra $h$, and let $\sigma_1,\sigma_2,\sigma_3$ be
elements of $g$ which commute with $\tau_j$ and which satisfy the $su(2)$ relations. A
 solution of the equations is
\begin{equation}
a=\tau_1+{\sigma_1\over 2s},~~b=\tau_2+{\sigma_2\over 2s},~~a=\tau_1+{\sigma_3\over 2s}.
\end{equation}
and this can be used as the boundary conditions for the non-regular triple.
The residue of the Higgs field is $\tau_2+i\tau_3+\sigma$, where $\sigma$
is the nilpotent element in the centralizer of $(\tau_1,\tau_2,\tau_3)$.

We have following theorem proved in \cite{kova}: Given $(\tau_2,\tau_3)$ and $\sigma$
is a nilpotent element which lives in the centralizer of $(\tau_2,\tau_3)$, the
moduli space of solutions with the above boundary conditions
is isomorphic to the adjoint orbit $\tau_2+i\tau_3+\sigma$, and it has a family
of hyper-Kahler structures. The family is parameterized by $\tau_1$ such that the
centralizer of the pair $(\tau_2, \tau_3)$ and the triple $(\tau_1,\tau_2,\tau_3)$
coincide. For the proof and more technical details, see \cite{kova}.

\end{document}